\documentclass[11pt,a4paper]{article}
\usepackage{microtype}
\usepackage[a4paper,margin=2.5cm]{geometry}
\usepackage{amsmath,amssymb}
\usepackage[T1]{fontenc}
\usepackage[utf8]{inputenc}
\usepackage{textcomp}
\usepackage{lmodern}
\usepackage{graphicx}
\usepackage{booktabs,tabularx,array,longtable}
\usepackage{enumitem}
\usepackage{float}
\usepackage[font=small,labelfont=bf]{caption}
\usepackage{tikz}
\usepackage[colorlinks=true,linkcolor=black,citecolor=black,urlcolor=blue!50!black]{hyperref}
\allowdisplaybreaks
\title{HEP-like 1550\,nm wide-field pixelated optical receiver for inter-satellite laser links: a simulation-based proof of concept for Earth, lunar and Earth-Moon scenarios}
\author{Maurizio De Santis$^{1}$\\[2pt] \normalsize $^{1}$Telespazio S.p.A.}
\date{}
\begin{document}
\maketitle

\begin{abstract}

I propose and evaluate by simulation a wide-field dome optical receiver (a field of view larger than a hemisphere) at 1550 nm for the acquisition of inter-satellite laser links. The detector is an InGaAs pixel hybrid of about 1 Mpixel with event-driven, time-stamped readout, inspired by the pixel detectors of high-energy physics. A pulsed beacon is detected by the coincidence of m hits in a sliding block of pixels, and the angle of arrival (AoA) is estimated by clustering and centroiding and handed to a narrow-field data channel. In a 550 km Earth constellation, with a 5 mrad beacon and a 10 mm aperture, detection takes 12 \textmu{}s to 2 ms (up to 116 ms with the 6--9 mm that the optics allow), with a whole-frame false-alarm rate below 0.01/s, and the AoA error is 0.13--0.23 mrad. Cooling to about 233 K gives a service life above 5 years for fluences up to about 2\ensuremath{\times}10\textsuperscript{9} p/cm\textsuperscript{2} per year. The receiver is also studied around the Moon and on the Earth-Moon link, where it works only between high orbits. Compared phase by phase with published scan-based acquisition, the search phase (3--18 ms with 10 mm) is shorter by a factor of about $10^{2}$ to $10^{5}$; end to end, with assumed pointing and loop-closure times, it takes 0.16--1.5 s, comparable to a terminal with calibrated open-loop pointing but without requiring that pointing at the receiver; the partner's beacon must still be pointed at it to about 1.25 mrad (5 mrad beacon). All values are simulation results, obtained with idealised optics and a 1 Mpixel detector that does not exist today; none is a measurement.

\end{abstract}

\section{Introduction}

Laser terminals on satellites acquire in seconds or tens of seconds (Section 2.1), either by scanning a cone of uncertainty or by searching for a beacon, with quadrant detectors or focal-plane arrays [32]. Low-orbit constellations multiply the links to be established and re-established, and future lunar infrastructure adds new distances and geometries. A wide-field acquisition sensor, which sees almost all of the useful sky and directly provides the angle of arrival (AoA), should reduce the scan and the dependence on accurate a-priori pointing.

My starting point comes from particle physics: the hybrid pixel detectors of ATLAS [14] and of the Medipix/Timepix family [15] read out each hit with its address, time stamp and amplitude (time over threshold, ToT); in reconstruction the hits are grouped into clusters and a position is computed from them. I ask whether this technical heritage, carried over to InGaAs at 1550 nm, is useful for inter-satellite acquisition.

The paper is a simulation-based proof of concept. I present no prototype; the assumptions are declared in the Appendix and every result depends on them.

\subsection{My contributions}

\begin{enumerate}[leftmargin=*, itemsep=2pt]
\item A receiver architecture with a dome receiver and a hybrid pixel detector for acquisition, and a narrow-field data channel. The idea I propose is the transfer of the design philosophy of HEP pixel detectors to the acquisition of inter-satellite links.
\item A model of pulsed-beacon detection by m-hit coincidence in sliding blocks, with the false-alarm rate (FAR) over the whole frame, and a Monte Carlo check of the formula for one block.
\item A pixel-level clustering and centroid model (binary with saturation correction, ToT), with gain, crosstalk, dead-pixel and jitter systematics.
\item The tolerance of per-hit attitude correction, a known technique, to synchronisation and scale errors in this receiver.
\item A system analysis: radiation damage and service life (with AP9 fluences from SPENVIS), thermal budget, and the Sun, Earth and Moon in the field of view (FOV).
\item Three scenarios: an Earth constellation (the strong case), a lunar constellation and the Earth-Moon link, each with its own geometric constraints.
\item A phase-by-phase comparison of the acquisition time with published scan-based terminals and with a terminal with calibrated open-loop pointing (Section 2.4).
\end{enumerate}

\subsection{Scope}

When the direction of the partner is known to within a few milliradians, a conventional narrow-field sensor on the data terminal is sufficient, and nothing in this paper argues against it. If, however, the geometry proposed here is adopted, the wide field becomes useful when the a-priori pointing is poor: after a loss of link, at start-up, or with a large attitude uncertainty. In those cases the receiver would find the partner and measure its direction without a scan. The partner's beacon must still be pointed at the receiver, to about a quarter of its divergence (about 1.25 mrad for a 5 mrad beacon, Section 5.1): the receiver removes the need for accurate pointing on the receiving side and relaxes the one at the transmitter, but it does not remove it. I therefore present the architecture as an acquisition aid for the data terminal.

\section{State of the art and positioning}

\subsection{Operational terminals}

Terminals publish acquisition times from 2 to 60 s and use scanning or a cone of uncertainty (Tesat: spatial acquisition within 2 s and frequency acquisition within 8 s in space-to-space links, and 22 s on NFIRE and 34 s on TerraSAR-X [11]; CONDOR about 30 s the first time and 2 s afterwards [23]; CrossBeam about 10 s after on-orbit calibration [24]; the requirement of the ESA Specification for Terabit/sec Optical Links (ESTOL) is below 60 s within the uncertainty cone [29]). Starlink, Kuiper and Telesat publish neither times nor sensors. For the LCRD/ILLUMA-T demonstration from geostationary orbit (GEO) to low Earth orbit (LEO), NASA reports a spiral acquisition scan of about 7 minutes (maximum radius 1 mrad), with 85\% of acquisitions within 5 minutes in June 2024 [12]. The scan can be avoided with an accurate open-loop pointing, and a few sources separate the search from the rest of the acquisition; they are used in Section 2.4 [34]--[38].

\subsection{Prior art identified in public sources}

The individual elements exist: omnidirectional receivers with AoA from discrete detectors [4], [3] a distributed photodiode array [5], with a cascaded coarse (photodiode array) and fine (quadrant detector) AoA measurement for wide-field pointing, acquisition and tracking (PAT) [39]; pixel arrays for detecting and tracking the partner [2]; event-driven readout for inter-satellite link (ISL) acquisition ([1], [6]); event-driven InGaAs sensors [8]; hybrid pixel cameras with time-stamping and cluster centroids in silicon [9]; an analysis of signal acquisition with photon-counting detector arrays in free-space optical links [28]; and Geiger-mode avalanche photodiode (APD) arrays in photon-counting receivers for deep-space communication [7]; coincidence processing is a standard technique of photon-counting systems, and I use it for beacon acquisition and do not claim it as new. I did not find the combination in the public literature. The search used only public sources (papers, proceedings, agency reports, datasheets and the open text of patents); I had no access to paywalled articles, proprietary documents or commercial patent databases, and I make no statement about their content.

\subsection{Positioning with respect to Raytheon BBN and MIT Lincoln Laboratory}

The comparison with the two sources closest to this work is summarised in Table 1. It is based on their publicly available texts (BBN [1]: the full text of the two patents on the USPTO site; MIT LL [2]: the public technology highlight and the text of the two related patents, US 10,581,521 and US 10,931,372, on Google Patents). It is not a legal reading of the claims.

\begin{table}[htbp]
\centering
\footnotesize
\caption{Positioning with respect to Raytheon BBN and MIT Lincoln Laboratory.}
\begin{tabularx}{\linewidth}{>{\raggedright\arraybackslash}X >{\raggedright\arraybackslash}X >{\raggedright\arraybackslash}X >{\raggedright\arraybackslash}X}
\toprule
 & \textbf{Raytheon BBN (US11063667, US11522612) [1]} & \textbf{MIT LL Dual-Mode Imaging Receiver [2]} & \textbf{This work} \\
\midrule
Detector & Event camera (per-pixel brightness change) behind a wide-angle lens; wavelength not specified & Single-photon detector array (256\ensuremath{\times}256 in the highlight; Geiger-mode APDs in the patent description) with dedicated readout integrated circuit (ROIC); pixels switch between photon counting and photon timing & InGaAs hybrid, about 1 Mpixel, 1550 nm \\
Detection & Events grouped by pixel proximity and time (about 1 ms in the example); unrelated events discarded as noise; no count threshold or false-alarm analysis & Per-pixel mode switching: counting pixels read synchronously, timing pixels asynchronously; high-pass temporal filters in the highlight; no coincidence between pixels claimed; detection in a fraction of a second & m-hit coincidence in sliding blocks, false alarms over the whole frame \\
Angle estimation & 3D location vector (azimuth and elevation) used to identify the source (independent claims of US 11,522,612); no centroid recited; accuracy not given & Transmitter centroid from the detecting photodetectors (dependent claim 20 of US 10,931,372); accuracy not given & Clustering and centroid (binary and ToT); 0.13--0.23 mrad (binary) and 0.10--0.65 mrad (ToT) simulated with realistic systematics \\
FOV & Wide-angle lens, no value given & Wide in the highlight; about 0.05 sr (about 14\textdegree{}) in the patent description & Dome, about 8.5 sr useful in LEO at 550 km \\
Application & Optical communication, including satellites (dependent claim 16 of US 11,063,667); demultiplexing of several simultaneous sources & Free-space optical communication; claims cover tracking a moving transmitter and receiving from several transmitters (claims 15 and 21 of US 10,931,372) & ISL: Earth, lunar and Earth-Moon \\
System aspects & Not covered; the location estimate uses camera position and orientation & Up to 10 simultaneous links & Radiation, thermal, attitude, Sun/Earth/Moon \\
\bottomrule
\end{tabularx}
\end{table}

\subsection{Indicative comparison of acquisition time}

I compare the acquisition time phase by phase, because the published figures cover different things. 

\subsubsection{Search phase only}

The scan of a cone of uncertainty takes from seconds to minutes: about 11--15 s to the first valid attitude information and 17--77 s to close the bidirectional tracking loop, for an uncertainty of $\pm$0.1\textdegree{} and a beam divergence of 193 \textmu{}rad, in a preliminary ground campaign with nine runs per pattern [35]; a mean of 131 s for a spiral scan with an uncertainty region of 3 mrad and a beam of 135 \textmu{}rad, in a simulation with the parameters of GRACE-FO [34]; about 7 minutes for LCRD/ILLUMA-T [12]; 2 s for the spatial acquisition of the Tesat terminals [11]. The scan time grows with the uncertainty [34], [36]. The search phase of my receiver is the light time plus the detection and the collection of the counts for a 0.3 mrad angle: 3.4, 11 and 18 ms at 1002, 3350 and 4910 km with a 10 mm aperture (3.4, 16 and 132 ms with 6 mm), dominated by the light time. This is shorter than the references by a factor of about $10^{2}$ to $10^{5}$ (10 mm; about 15 to $10^{5}$ with 6 mm), and it does not depend on the initial pointing uncertainty of the receiver (the beacon of the partner must still be pointed at it, Section 5.1). The comparison is indicative: the 5 mrad beacon is wider than the 135--193 \textmu{}rad of those terminals, the uncertainties and ranges differ, and my value is a simulation with an ideal filter and a dead time of 1 \textmu{}s, at the optimistic end of the range in the literature. 

\subsubsection{End to end}

After the search, both approaches need the same remaining steps: pointing of the narrow channel, closing the tracking loop and the exchange of signals over one or two light times. I call $S$ the time of these steps and I estimate it as follows. Pointing, 0.1--1 s: the lower end is the 16 ms per step of a fast steering mirror reported in [36]; the upper end is the time to move such a mirror by a few milliradians at the rates of 5--8.5 mrad/s used in a recent simulation of pointing delays [37]. Loop closure, 0.05--0.5 s: the track sensor is sampled at 1 kHz [37], so this is 50--500 samples, a number that [37] does not give, and a survey states that the transition from acquisition to tracking typically takes less than one second [38]. Light times of the exchanges that follow the search: 3--33 ms from the geometry (the first one-way light time is already in the search time of the receiver). These are order-of-magnitude estimates of mine, not measurements on this architecture. They also leave out the coarse slew towards the partner, which both approaches need and which a simulation puts at between several seconds and tens of seconds [37]. I therefore give the result as a function of $S$ (Table 2): with $S$ = 0.15, 1.5 and 10 s my end-to-end time is 0.16, 1.5 and 10 s, and the advantage over the scan-based references goes from about 1.2 (Tesat, $S$ = 10 s) to about 2600 (LCRD/ILLUMA-T, $S$ = 0.15 s). It assumes a receiver on both terminals and rests on a simulation, without measurement.

\begin{table}[htbp]
\centering
\footnotesize
\caption{End-to-end advantage of the receiver over scan-based acquisition as a function of the common time $S$ (pointing of the narrow channel, loop closure and light times) that both approaches need. The receiver's search time is 11 ms (median case, 10 mm). Each ratio is (scan time + $S$) / ($S$ + 11 ms).}
\begin{tabularx}{\linewidth}{>{\raggedright\arraybackslash}X >{\raggedright\arraybackslash}X >{\raggedright\arraybackslash}X >{\raggedright\arraybackslash}X}
\toprule
\textbf{Reference (scan time)} & \textbf{$S$ = 0.15 s} & \textbf{$S$ = 1.5 s} & \textbf{$S$ = 10 s} \\
\midrule
My end-to-end time & 0.16 s & 1.5 s & 10 s \\
Tesat, spatial acquisition (2 s) [11] & 13 & 2.3 & 1.2 \\
DLR, first attitude information, $\pm$0.1\textdegree{} (13 s) [35] & 82 & 9.6 & 2.3 \\
Spiral scan with GRACE-FO parameters (131 s) [34] & 810 & 88 & 14 \\
LCRD/ILLUMA-T spiral scan (420 s) [12] & 2600 & 280 & 43 \\
\bottomrule
\end{tabularx}
\end{table}

\noindent

\subsubsection{Terminals that do not scan}

A scan can also be avoided with an accurate open-loop pointing. A ground-based single-terminal test reports an equivalent acquisition time of 0.91 s on average, with every run under 1 s, for a terminal with calibrated pointing (mean error 120 \textmu{}rad) [36]; the counterpart is emulated by a star, the slew of the gimbal is subtracted and the propagation over 45,000 km is computed, so it is an equivalent time and not an in-orbit measurement. Its 0.91 s is a total that includes its own steering-mirror, processing and propagation terms, so it is to be compared with my $S$ + 11 ms (0.16--10 s) and not with the search time: with $S$ = 0.15 s I would be faster and with $S$ = 1.5 s slower, so I claim no shorter time. The difference is in what is required: that approach needs the accurate a-priori pointing, obtained by a calibration with a star camera (an open-loop uncertainty of 5--20 mrad is quoted for conventional terminals [36]), whereas the receiver proposed here measures the direction of the partner and does not depend on the pointing of its own terminal; it does need the beacon of the partner to be pointed at it, to about 1.25 mrad (1$\sigma$ per axis) for a 5 mrad beacon (Section 5.1), which is between the 5--20 mrad quoted for conventional terminals and the 0.12 mrad of the calibrated one. The case for the receiver is therefore a time of the same order without the calibrated pointing and the scan of the receiving terminal, after a loss of link, at start-up or with a large attitude uncertainty of the receiver. The two systems also solve different problems: the terminals close a communication link, while my receiver finds the partner and measures its direction, and the closing of the link remains with the narrow channel.

\subsection{Technical heritage}

The methods of event-driven readout, clustering and ToT derive from high-energy-physics (HEP) hybrid detectors, and I adopt them from that field. The pixel-level ideas are not new: event-driven readout, time-stamping, ToT clustering and centroids appear in the sources above or in HEP detectors. Three further elements are also known: omnidirectional coverage with discrete detectors, full-sky in the claims of [3] (see also [4]); the grouping of events by adjacent pixels and time in [1]; and the correction of each event for the rotation of the camera according to its time-stamp, used in event-based vision (for example [33]). As far as I could verify with public sources, I did not find these elements combined, and I propose as new only that combination and its quantification: (i) a single pixelated receiver (about 1 Mpixel, 2.9 mrad per pixel) covering more than a hemisphere, about 8.5 sr useful, whereas [3], [4] use discrete detectors, and [1] describes a wide-angle lens on a focal plane array without quantifying the field or the array size; (ii) a detection threshold of m hits in sliding blocks, with the false-alarm rate computed over the whole frame of a large array. Thresholds set from a desired false-alarm probability are known for photon-counting detector arrays [28], and [1] filters events by space-time density with no stated count threshold or false-alarm rate, so what is mine here is the block-wise form on a large array and its whole-frame false-alarm budget; (iii) the quantification of the tolerance of per-hit attitude correction, a known technique, to synchronisation and scale errors for this receiver; and (iv) a system-level evaluation (geometry, radiation, cooling, attitude, Sun/Moon/Earth) for Earth, lunar and Earth-Moon links. The first two are the architectural content and the last two are analysis. The thread that ties these elements together is the design philosophy of high-energy-physics pixel detectors (readout of each hit with address, time stamp and ToT, coincidence, cluster reconstruction), which, as far as I could verify, is transferred here as the design principle of an inter-satellite acquisition receiver; the individual ingredients have precedents, as stated above. These are simulation results, not measurements.

\section{Proposed architecture}

The architecture has two channels: a dome receiver for acquisition, which finds the beacon and estimates the AoA, and a narrow-field data channel that receives the direction and closes the link.

\begin{enumerate}[leftmargin=*, itemsep=2pt]
\item \textbf{Optics.} Dome with pixel or pixel-group optics. The useful field is about 8.5 sr in LEO at 550 km, which is more than a hemisphere ($2\pi$ = 6.28 sr; up to about 111\textdegree{} from zenith) because partner satellites lie near or below the local horizon. The Earth disc (half-angle 67.0\textdegree{}) is excluded and a margin towards the limb (tangent at 100 km: 69.2\textdegree{} from nadir). A pixel subtends 2.92 mrad at 1 Mpixel.
\item \textbf{Detector.} InGaAs hybrid of about 1 Mpixel, 25 \textmu{}m pitch (assumption), with event-driven, time-stamped readout. Two modes considered: binary photon counting with single-photon avalanche diodes (SPAD) or linear with APD, ROIC and ToT.
\item \textbf{Beacon detection.} Coincidence of at least m hits in a sliding block of 2\ensuremath{\times}2 or 3\ensuremath{\times}3 pixels within a window W. A pixel records at most one hit per pulse (dead time 1 \textmu{}s).
\item \textbf{Clustering and AoA.} After the trigger a 9\ensuremath{\times}9 region of interest (ROI) is read out, weighted by the hit rate (corrected for saturation) or by ToT, and the centroid is computed.
\item \textbf{Masks.} Hot or damaged pixels, and the regions of the Sun, Moon and Earth disc, are masked using ephemerides and star maps.
\item \textbf{Attitude.} Hit time-stamps are aligned with the gyro and star tracker to correct the jitter hit by hit.
\item \textbf{Handover.} An AoA with an error below about 0.3 mrad takes the narrow channel directly onto the beacon.
\end{enumerate}

Figure 1 summarises the chain, from the beacon to the handover. Figure 2 is a conceptual drawing of the receiver: the dome with the FOV tiled in pixels, and the cross-section of one acquisition channel with optics, filter, hybrid sensor, readout, cooler and handover to the narrow-field channel.

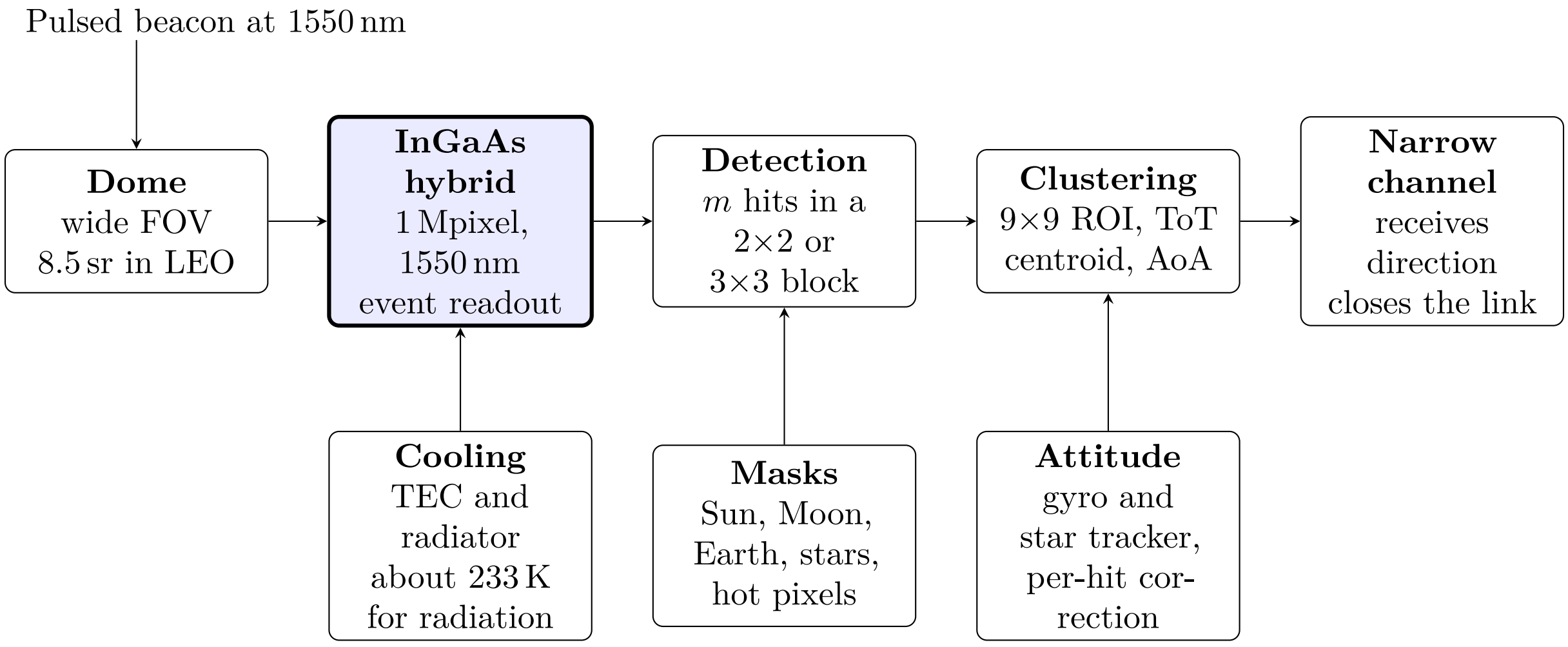
\begin{figure}[htbp]
\centering
\resizebox{\linewidth}{!}{%
\begin{tikzpicture}[font=\small, >=stealth,
  blk/.style={draw, rounded corners=3pt, minimum width=2.7cm, minimum height=1.5cm, align=center, text width=2.5cm},
  hi/.style={blk, very thick, fill=blue!8}]
\node[blk] (dome) at (0,0) {\textbf{Dome}\\wide FOV\\8.5\,sr in LEO};
\node[hi] (det) at (3.4,0) {\textbf{InGaAs hybrid}\\1\,Mpixel, 1550\,nm\\event readout};
\node[blk] (trg) at (6.8,0) {\textbf{Detection}\\$m$ hits in a\\$2{\times}2$ or $3{\times}3$ block};
\node[blk] (clu) at (10.2,0) {\textbf{Clustering}\\$9{\times}9$ ROI, ToT\\centroid, AoA};
\node[blk] (nar) at (13.6,0) {\textbf{Narrow channel}\\receives direction\\closes the link};
\node[blk, minimum height=1.9cm] (cool) at (3.4,-3.3) {\textbf{Cooling}\\TEC and radiator\\about 233\,K\\for radiation};
\node[blk, minimum height=1.9cm] (msk) at (6.8,-3.3) {\textbf{Masks}\\Sun, Moon,\\Earth, stars,\\hot pixels};
\node[blk, minimum height=1.9cm] (att) at (10.2,-3.3) {\textbf{Attitude}\\gyro and\\star tracker,\\per-hit correction};
\draw[->] (dome) -- (det); \draw[->] (det) -- (trg); \draw[->] (trg) -- (clu); \draw[->] (clu) -- (nar);
\draw[->] (cool) -- (det); \draw[->] (msk) -- (trg); \draw[->] (att) -- (clu);
\draw[->] (0,1.9) -- (dome.north);
\node[anchor=south west] at (-1.3,1.85) {\small Pulsed beacon at 1550\,nm};
\end{tikzpicture}}
\caption{Two-channel architecture: five signal blocks and three support blocks. The highlighted block is the event-driven hybrid detector, inherited from HEP pixel detectors. The cooling block combines a thermoelectric cooler (TEC) and a radiator.}
\label{fig:arch}
\end{figure}

\begin{figure}[htbp]
\centering
\includegraphics[width=\linewidth]{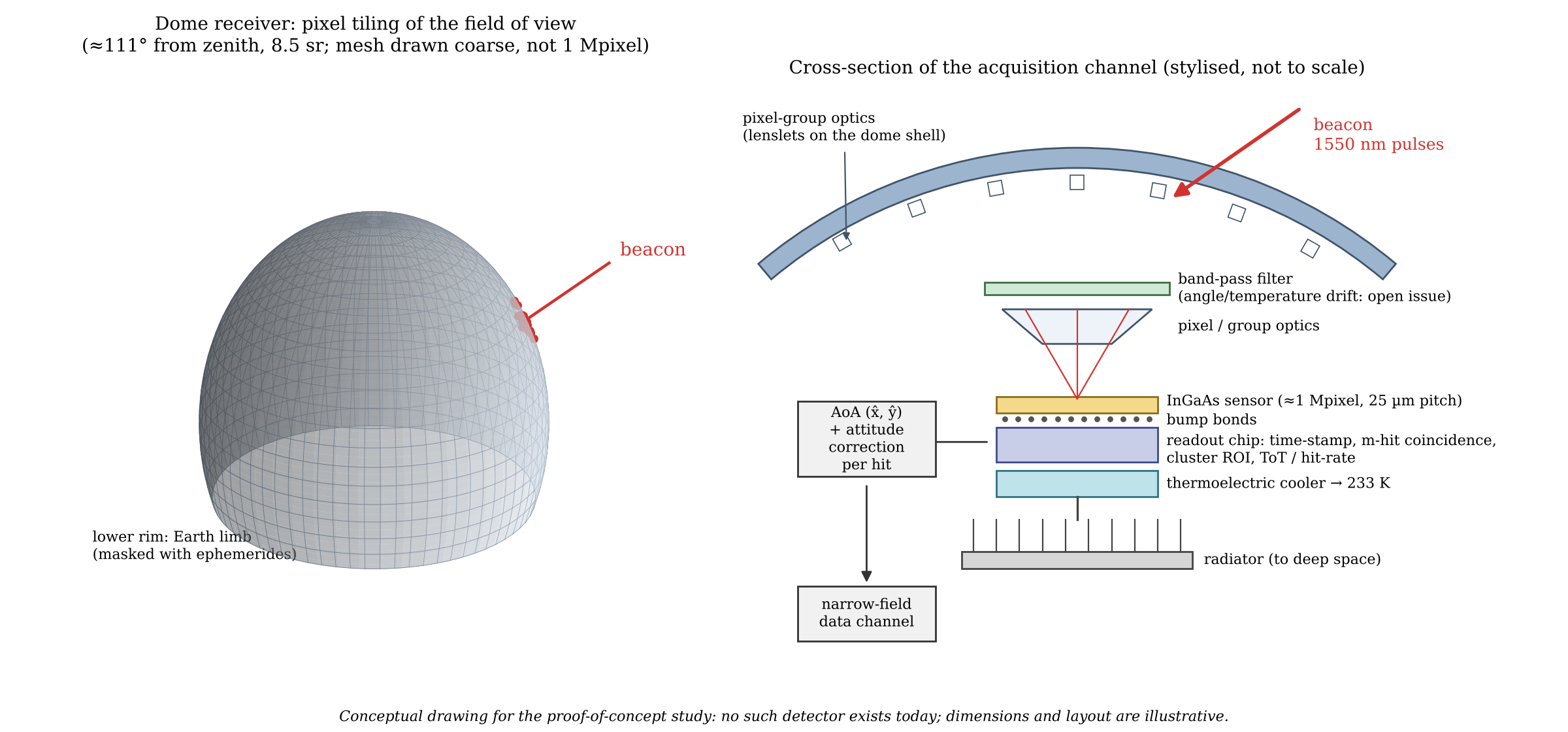}
\caption{Conceptual drawing of the receiver. Left: the dome with the FOV tiled in pixels (about 111\textdegree{} from zenith, 8.5 sr; the mesh is drawn coarse). Right: cross-section of one acquisition channel, from the optics of a pixel group (the small squares on the dome shell) and the band-pass filter to the hybrid InGaAs sensor, the readout chip, the cooler and radiator, and the handover to the narrow-field channel. Schematic drawing, out of scale.}
\end{figure}

\subsection{Why the HEP pixel philosophy transfers}

Beacon acquisition and HEP triggering share the same structure: a rare, short signal must be separated from a dense and uncorrelated background, over a large number of channels, with a false-trigger budget defined for the whole detector. HEP pixel systems address this with a few design choices that apply directly here:

\begin{enumerate}[leftmargin=*, itemsep=2pt]
\item \textbf{A rate budget instead of a per-pixel threshold.} A HEP trigger is specified by the accepted background rate over the whole detector. The detection criterion adopted here does the same: the coincidence parameters (m, W, block size) are chosen so that the whole-frame FAR stays below 0.01/s.
\item \textbf{Local coincidence before readout.} At the dark-count level assumed here a 1 Mpixel array produces of the order of 10\textsuperscript{10} hits/s, far above what a readout can carry. HEP detectors face the same problem and reduce the data at the front end with local trigger logic. The block coincidence is the equivalent step; it is assumed to be implemented on the detector, and its electronics are left out of scope.
\item \textbf{Time-stamped, event-driven data.} Hits carry their time of arrival (and, in the linear mode, their ToT). Time stamps are what make coincidence, the attitude correction hit by hit and the rejection of uncorrelated noise possible.
\item \textbf{Clusters and centroids.} Timepix-type detectors [9], [15] use the charge or ToT of a cluster to place the impact point to a fraction of a pixel. Applying the same idea to the beacon spot is what, in the simulation, brings the AoA error to 0.13--0.23 mrad with the binary estimator and 0.10--0.65 mrad with ToT (realistic systematics), with a pixel of 2.92 mrad; the result depends on the spot-size and jitter assumptions stated in Sections 4.4 and 4.8.
\item \textbf{Hybrid construction.} The sensor and the readout chip are separate and bump-bonded, so the sensor material can be changed without redesigning the readout concept. This makes a move from silicon to InGaAs at 1550 nm conceivable; it is also why the missing piece today is a 1 Mpixel InGaAs hybrid. A Geiger-mode InGaAs pixel is a SPAD with its own dead time, afterpulsing and dark-count behaviour, and the ToT-like linear mode assumes an APD chain separate from the SPAD chain of the binary mode.
\item \textbf{Calibration and masking as part of the design.} HEP pixel systems are operated cooled, calibrated pixel by pixel, and run with a fraction of masked or noisy channels, including after displacement damage. The systematics model (gain, crosstalk, dead pixels), the hot-pixel masking and the cooling strategy follow the same practice.
\end{enumerate}

The transfer concerns a design philosophy and a set of well-tested methods. It is a working hypothesis, supported by the simulations and to be confirmed by measurement. The space environment, the power and mass limits and the absence of a 1 Mpixel InGaAs hybrid are the real differences, and they are treated as limits (Section 8).

\subsection{Optical constraints}

The FOV and the pixel size fix the optics. The focal length is about 8.6 mm (pitch divided by pixel angle), so the aperture of the lens of one pixel group is about 6 mm with the fastest commercial SWIR lenses (f/1.4) and up to about 9 mm only if an f/1 design were available (assumption). I use 10 mm (f/0.86) as the nominal aperture of the simulations, which is slightly above these values and is therefore optimistic. With 9 mm the detection times at 5 mrad are 14 \textmu{}s, 0.36 ms and 4.3 ms in the near, median and far cases, and with 6 mm they are 33 \textmu{}s, 5.1 ms and 116 ms; the far case with the smallest realistic aperture is therefore almost two orders of magnitude slower than with 10 mm (2.0 ms). A larger collecting area would need several apertures or a micro-lens array. I also keep a 30 mm aperture, and in the Earth-Moon link apertures up to 100 mm, as hypothetical reference cases to show how much a larger collecting area would give; they are ideas, not baselines. A 30 mm aperture on a pixel of 2.92 mrad and a 25 \textmu{}m pitch would be f/0.29, which is not realisable with a single lens per pixel group. It is conceivable as a design idea, for example an f/1 lens of 30 mm focal length, but at the same angular pixel the pitch would grow to about 90 \textmu{}m and a 1 Mpixel detector to about 90 mm on a side (my estimate from the geometry); it would cost much more and weigh much more. In addition, a narrow band-pass filter of 1 nm shifts towards shorter wavelengths with the angle of incidence and drifts with temperature, so over a wide field a filter of that width is not guaranteed to transmit the beacon; the background and signal figures in this paper assume an ideal filter, and this is an open design issue.

The dome is the mechanical surface that carries the pixel or pixel-group optics; it is not itself a lens. Its material (for example fused silica for any window part, with anti-reflection coatings at 1550 nm), thickness, mounting and radiation hardening are assumptions that are not optimised here, and the dome-filter reflections are not modelled (Section 8).

\section{Models and methods}

All parameters are assumptions declared in the Appendix.

\subsection{Geometry}

Constellations are described by the Walker Delta notation i:t/p/f [16] (inclination, total satellites, orbital planes, phasing factor; the phase offset between adjacent planes is f\ensuremath{\times}360\textdegree{}/t). The Earth case is a Walker 53\textdegree{}:1584/72/1 at 550 km, i.e. 72 planes of 22 satellites with a phase offset of about 0.23\textdegree{} between adjacent planes. It has the same inclination, altitude and plane and satellite counts as the first published Starlink shell; the phasing factor f = 1 is my assumption, since the operator's actual value is not public. The lunar case is a Walker 65\textdegree{}:24/4/1 (4 planes of 6 satellites) at 1000 and 5000 km lunar altitude. Three cases for each constellation: near (5th percentile of distances), median, and far (maximum). In the Earth case: 1002, 3350 and 4910 km, computed at one epoch over the pairs, with line of sight above 100 km altitude, in which the first satellite is one of every eighth (using all satellites the 5th percentile is 1005 km; the median and the maximum do not change), including the fixed ISL neighbours (whose in-plane spacing is about 1970 km). Earth-Moon: 356,500, 384,400 and 406,700 km between centres; the near and far cases subtract and add the orbital radii of a 550 km Earth orbit and of a 1000 km lunar orbit (6921 and 2737 km), giving 346,842 and 416,358 km.

\subsection{Detection}

Pulsed beacon with 0.5 W average power, 20 \textmu{}J maximum energy per pulse and 8 ns duration (2.5 kW peak). The coincidence window $W$ cannot be shorter than the pulse, so it is searched among 10, 20 and 50 ns. The coincidence start rate in the Poisson background and the whole-frame FAR are:

\begin{gather}
R_{start} = \lambda_b \, P\left[N \ge m-1\right] \notag\\
N \sim \mathrm{Poisson}(\lambda_b W) \notag\\
\mathrm{FAR} = N_{pix}\, R_{start} \le 0.01\ \mathrm{s}^{-1}
\end{gather}

I consider 2\ensuremath{\times}2 and 3\ensuremath{\times}3 blocks, with m, W and the period Tp optimised for each case. The Monte Carlo generates the Poisson dark-count stream of a single block and counts the coincidence starts (a start is a hit that finds at least m$-$1 further hits within W); it is compared with the analytical rate of one block. To collect enough events I used an inflated dark-count rate of 2\ensuremath{\times}10\textsuperscript{5}/s per pixel (20 times the nominal value) and windows of 100 and 300 ns, longer than the design windows of 10--50 ns. Four configurations were tested: 3\ensuremath{\times}3 blocks with (m, W) = (3, 100 ns) and (4, 300 ns), and 2\ensuremath{\times}2 blocks with (2, 100 ns) and (3, 300 ns). Each is the mean and standard error (SE) of 100 runs of 1 s, with 2.0\ensuremath{\times}10\textsuperscript{6} to 6.2\ensuremath{\times}10\textsuperscript{6} starts per configuration. The ratios to the analytical rate are 1.0003 $\pm$ 0.0008, 0.9995 $\pm$ 0.0007, 1.0004 $\pm$ 0.0004 and 1.0002 $\pm$ 0.0007, consistent with the Poisson error of the counts (0.04--0.07\%). This checks the algebra of the formula for one block at a high rate; it does not test the operating false-alarm rate of 0.01/s, which is obtained analytically, nor correlated noise. The formula assumes uncorrelated dark counts only (Section 8).

\subsection{Choice of the nominal dark-count rate (DCR)}

The value $\mathrm{DCR} = 10^{4}$/s per pixel at 233\,K is a design assumption: it is the central value of the $10^{3}$--$10^{5}$/s range that I vary in the results. As a plausibility check, the largest published InGaAs/InP SPAD array that I found ($96\times96$ pixels, 25\,\textmu{}m pitch, 15\% photon detection efficiency (PDE) [20]) has a median DCR of 87 kcps at room temperature (295\,K, gated mode, median over a $6\times24$ pixel subset, range 45--175 kcps). If the dark count halves every 8\,K (the same assumption used for the radiation model), cooling from about 296\,K to 233\,K lowers it by a factor $2^{63/8} \approx 235$, to about 370/s. The nominal value is therefore about 27 times higher than this extrapolation, which leaves a margin for effects that I do not model: the higher bias needed for a PDE of 0.30, a different pixel pitch, and a halving temperature different from 8\,K (Section 5.4 shows that 6\,K or 10\,K changes the radiation life by an order of magnitude). The extrapolation is not a validation. With the nominal 10\,mm aperture the sensitivity to this choice grows with distance: the detection time at $10^{3}$, $10^{4}$ and $10^{5}$/s is 11.6, 11.8 and 13 \textmu{}s in the near case, 0.13, 0.20 and 1.6 ms in the median and 0.58, 2.0 and 65 ms in the far case, because the weaker signal is closer to the background. With the hypothetical 30\,mm aperture the time changes by less than 15\% over the same range (1.3, 14 and 31 \textmu{}s at $10^{4}$/s).

\subsection{Clustering}

Per-pixel simulation in a 9\ensuremath{\times}9 ROI with a defocused point spread function (PSF) (0.4--1.2 pixel in the clustering study; 0.4, 0.6 and 0.8 pixel in the systematics runs that give the values quoted). Three estimators: B1 binary, B2 binary with saturation correction (weight $-$ln(1$-$f)), and L linear with APD (G=20, F=9.2, $\sigma$e=50 e\textsuperscript{$-$}, 5$\sigma$ threshold) and ToT. Systematics: per-pixel gain, crosstalk, dead pixels, jitter.

\subsection{Radiation}

Displacement damage with the constant anchored to a measurement on InGaAs PIN photodiodes (Hamamatsu, 105 MeV: from about 2--5 nA to 96 nA at 2.55\ensuremath{\times}10\textsuperscript{11} p/cm\textsuperscript{2} [10]; the source notes that its pre-irradiation values are affected by measurement noise). Cooling with the dark count halving every 8 K (assumption), a lognormal distribution of hot pixels, and masking above 20\ensuremath{\times} the nominal DCR.

\subsection{Thermal}

Node-based budget (dome, radiator, detector), with a TEC or a passive link, for beta of 0, 30, 60 and 76\textdegree{}.

\subsection{Background}

Sun (AM0 at 1550 nm), Moon, Earth with short-wave infrared (SWIR) reflectance 0.05--0.5, stars, zodiacal light and Harvey-type stray light.

\subsection{Attitude}

Correlated Gaussian jitter (10--1000 Hz band); attitude measured with 0.05 mrad white noise plus 0.05 mrad bias; per-hit correction; synchronisation and scale errors varied separately.

\subsection{Statistical uncertainty}

The results of this paper are partly Monte Carlo. Those on the AoA, on the false-alarm check and on the hot-pixel sample are Monte Carlo, that is, random sampling of the model described here; the others are analytical or deterministic calculations. The Monte Carlo runs sample my model under its stated assumptions (Poisson photon arrivals, independent pixels, at most one hit per pixel per pulse, Gaussian PSF), so they are numerical experiments on the model. Detection probabilities, thresholds, link budgets, orbital geometry and the thermal balance are analytical and have no sampling noise. The Monte Carlo parts are the AoA simulations (Sections 5.2, 5.3, 6.1 and 7.1), the check of the false-alarm formula (Section 4.2), the detection check on the pixel-level simulation (Section 5.2, 2\ensuremath{\times}10\textsuperscript{4} trials per spot size, SE of the probabilities at most 0.4\%) and the lognormal sample of hot pixels of the radiation model (Section 5.4). For the AoA I use $N$ = 3\ensuremath{\times}10\textsuperscript{4} trials per configuration, run as 30 independent batches of 1000 trials with fixed seeds. The reported error is the rms over the two axes of the whole sample. Its SE comes from the spread of the squared error between batches, SE = $s_{\mathrm{MSE}}/(2\sigma\sqrt{n_b})$, with $n_b$ = 30 and $s_{\mathrm{MSE}}$ the standard deviation of the batch means of the squared error, which does not assume Gaussian errors; for Gaussian errors it would be about $\sigma/(2\sqrt{N})$, i.e. 0.3\%. The ratio of ToT to binary error uses the propagation of the two relative SE; the thresholds of the tolerance study (Section 5.3) use a bootstrap over the batches (400 resamples). Probabilities are averaged over the batches.

The defocus $\sigma_{PSF}$ is chosen as the minimum over a grid of three values, and a minimum taken over noisy estimates is biased. I measured this on the median case with realistic systematics: single runs with 1000 trials per $\sigma_{PSF}$ scatter by 2.3\% (binary) and 2.4\% (ToT) rms around the 3\ensuremath{\times}10\textsuperscript{4}-trial value, with mean offsets of $-$0.1\% and +0.3\% (30 repetitions). With 3\ensuremath{\times}10\textsuperscript{4} trials the SE is 0.2--0.7\% (binary) and 0.2--1.2\% (ToT) and the selection effect is expected to be smaller than that. The check of the false-alarm formula is repeated 100 times for 1 s each, and the SE is the spread between repetitions, since the events within one run are correlated. The hot-pixel sample (4\ensuremath{\times}10\textsuperscript{5} pixels) changes the radiation lifetimes by 0.24\% rms between 20 independent samples (for example 8.06 $\pm$ 0.02 years at 233 K and 1.5\ensuremath{\times}10\textsuperscript{9} p/cm\textsuperscript{2}/year).

They quantify the sampling noise of the simulation; the validity of the model and of its assumptions is the subject of the Appendix and of Section 8.

\section{Main case: Earth constellation (the strong case)}

\subsection{Beacon detection}

The quantities plotted in Fig. 3 follow from the link budget. The beacon gives a mean rate of detected counts $r_{s}$, from the free-space link-budget (range) equation for a diverging beam, with the Planck--Einstein photon energy h$\nu$, and, per pulse, a mean number of counts $\mu$:

\begin{gather}
r_s = \frac{P_{tx}\,\eta_{opt}\,\mathrm{PDE}}{h\nu}\,\frac{d^{2}}{(\theta R)^{2}} \notag\\
\mu = C\, r_s\, T_p
\end{gather}

A pulse is detected when the signal counts s (spread over the N pixels of the block) plus the background counts reach the threshold m. The first factor is the binomial distribution of the signal over the block, the second the Poisson distribution of the background, and $p = 1 - e^{-\mu/N}$ is the Poisson probability that a pixel records at least one count:

\begin{gather}
P_d = \sum_{s=0}^{N} \binom{N}{s}\, p^{s}(1-p)^{N-s}\; \Pr\!\left[\mathrm{Poisson}(\lambda_b W) \ge m-s\right] \notag\\
p = 1-e^{-\mu/N} \notag\\
\lambda_b = N\,\mathrm{DCR}
\end{gather}

The time to detect with 95\% confidence (from the geometric distribution of independent pulses: $1 - (1 - P_{d})^{n_{p}} \geq 0.95$), to collect enough counts for the angle (from the $1/\sqrt{N}$ law below), and the acquisition time plotted in Fig. 3, are:

\begin{gather}
n_p = \max\!\left(1,\ \frac{\ln 20}{-\ln(1-P_d)}\right) \notag\\
t_{95} = n_p\,T_p \notag\\
n_{ang} = \left(\frac{\sigma_{PSF}}{\sigma_{target}}\right)^{2} \notag\\
t_{ang} = \max\!\left(1,\ \left\lceil \frac{n_{ang}}{\mu} \right\rceil\right) T_p \notag\\
t_{acq} = \max\left(t_{95},\, t_{ang}\right)
\end{gather}

where:

\begin{itemize}[leftmargin=*, itemsep=2pt]
\item $P_{\mathrm{tx}}$ = 0.5 W is the average beacon power, $\eta_{\mathrm{opt}}$ = 0.5 the optical efficiency, PDE = 0.30, h$\nu$ the photon energy at 1550 nm (about 1.28\ensuremath{\times}10\textsuperscript{$-$19} J);
\item d is the receiver aperture, $\theta$ the full beacon divergence, R the range, so that d\textsuperscript{2}/($\theta$R)\textsuperscript{2} is the fraction of the beacon spot collected by the aperture;
\item C = 0.7 is the average fraction of the signal captured by the sliding block, $T_{p}$ the pulse period, limited by the pulse energy $P_{\mathrm{tx}}$\ensuremath{\cdot}$T_{p}$ \ensuremath{\leq} 20 \textmu{}J;
\item N = 4 or 9 is the number of pixels in the block (2\ensuremath{\times}2 or 3\ensuremath{\times}3), m the coincidence threshold, W the coincidence window, DCR the rate per pixel and $\lambda_{b}$ the background rate on the block;
\item $\sigma_{\mathrm{PSF}}$ = 1.3 mrad is the spot size and $\sigma_{\mathrm{target}}$ = 0.3 mrad the required AoA precision, so that $n_{\mathrm{ang}}$ \ensuremath{\approx} 19 counts are needed for the angle.
\end{itemize}

For each case I optimise the architecture (2\ensuremath{\times}2 or 3\ensuremath{\times}3), m, W and $T_{p}$ to minimise $t_{\mathrm{acq}}$, under the whole-frame constraint FAR \ensuremath{\leq} 0.01/s of Section 4.2. Since $r_{s}$ scales as d\textsuperscript{2}/($\theta$R)\textsuperscript{2}, the strong dependence of Fig. 3 on range and divergence is, in the count-limited regime, a power law with exponent $-$2 in each; at large divergence the pulse period reaches its limit and the detection probability falls faster, as the 20 mrad bars show.

With a 5 mrad beacon, a 10 mm aperture and a DCR of 10\textsuperscript{4}/s per pixel, detection takes 12 \textmu{}s in the near case, 0.20 ms in the median and 2.0 ms in the far case, always with false alarms below 0.01/s over the whole frame. With a 0.75 mrad beacon the times drop to 1 \textmu{}s (the grid limit), 2.9 and 6.4 \textmu{}s; with 20 mrad they rise to 0.55 ms, 11 s and 7 minutes. With the hypothetical 30 mm aperture the 5 mrad times would be 1.3, 14 and 31 \textmu{}s, and the 20 mrad ones 21 \textmu{}s, 1.1 ms and 18 ms. A reference scan that I computed (a 0.75 mrad beam over a cone of $\pm$0.5\textdegree{}, about 1083 positions with an overlap of 2, each costing a round-trip light time plus 1 ms) takes 8--37 s, and real terminals publish 2--60 s for the whole pointing phase. My number covers detection and the time to collect the roughly 19 counts needed for a 0.3 mrad angle estimate; it excludes light time and the slew and settling of the terminal.

\textbf{Transmit pointing of the beacon.} The times above assume that the receiver is inside the beam of the partner's beacon, so the partner must point the beacon at it. With the convention of Section 7.1 ($\sigma_{\mathrm{pt}} \simeq \theta/4$ per axis), the allowed pointing error is 1.25 mrad ($1\sigma$ per axis) for a 5 mrad beacon, 2.5 mrad for 10 mrad and 5 mrad for 20 mrad; for the 5 mrad beacon the half-width of the beam, 2.5 mrad, is exceeded 14\% of the time. The error has two parts. The ephemeris part is the error in the relative position of the two satellites divided by the range: 1 km gives 1.0, 0.30 and 0.20 mrad at 1002, 3350 and 4910 km, and 100 m gives 0.10, 0.03 and 0.02 mrad. The attitude part is the pointing error of the transmitting terminal. Added in quadrature, a 1 km position error at 1002 km leaves 0.75 mrad for the attitude, whereas with 100 m or better almost the whole 1.25 mrad is available to it; a body pointing of the order of 0.05\textdegree{} (0.9 mrad), which is what a star-tracker-based attitude system can be expected to provide, would then be enough. These position and attitude errors are orders of magnitude that I assume, not values taken from a mission. The point-ahead angle is at most about 50 \textmu{}rad for two LEO satellites with aligned velocities (the sum of the transverse speeds, 2 $\times$ 7.59 km/s, divided by c), about 4\% of the budget, and it is deterministic. Compared with Section 2.4, 1.25 mrad is 4 to 16 times tighter than the 5--20 mrad of open-loop uncertainty quoted for conventional terminals [36], and about ten times looser than the 0.12 mrad mean error of the calibrated terminal of [36]; the same convention applied to the narrow beams of the references (135--193 \textmu{}rad) gives 34--48 \textmu{}rad. The receiver therefore does not remove the need for accurate pointing: it removes the need to point the receiving terminal and the scan, and it moves the requirement at the transmitter from a few tens of microradians to about a milliradian. A transmitter with a pointing error of 5--20 mrad would, if that were a $1\sigma$ error, need a beacon of 20--80 mrad; for 20 mrad the detection takes 0.55 ms, 11 s and 7 minutes with 10 mm (21 \textmu{}s, 1.1 ms and 18 ms with the hypothetical 30 mm), and 80 mrad is weaker still and I have not simulated it. Outside the beam the signal falls and the detection time rises steeply; the model treats the spot as uniform and does not quantify this.

\begin{figure}[htbp]
\centering
\includegraphics[width=\linewidth]{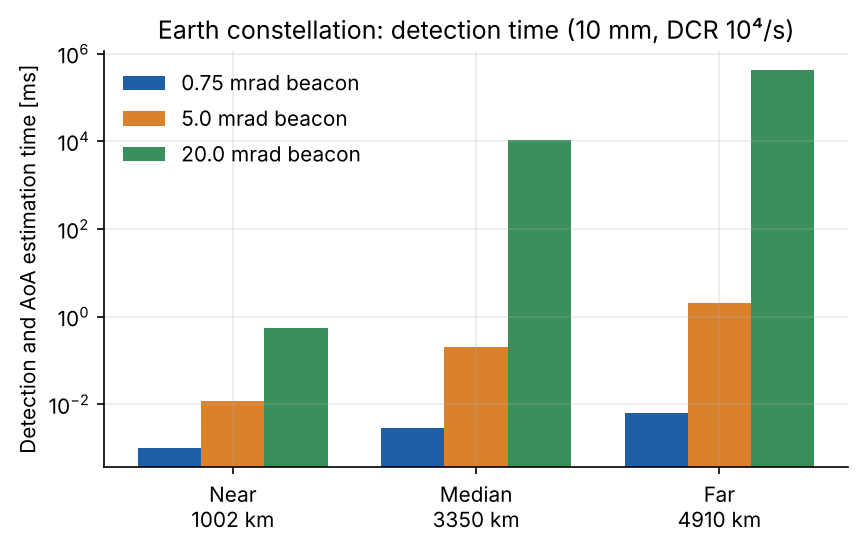}
\caption{detection time and AoA estimate, three cases and three divergences (10 mm, DCR 10\textsuperscript{4}/s). For 0.75 mrad the near-case value is the grid minimum of Tp.}
\end{figure}

\subsection{AoA accuracy and systematics}

Detection (Section 5.1) and angle estimation use the same link budget (5 mrad beacon, 10 mm aperture, same dark-count rate and dead time), but two different counts. The detection model keeps 70\% of the signal (C), the part falling in the sliding block, whereas the AoA simulation uses all the counts in the 9\ensuremath{\times}9 ROI. The AoA is estimated after detection, so the total time from the first pulse is the detection time plus the 0.8 ms of the 20 pulses. To check them in one model, I applied the detection rule of Section 5.1 (at least m hits in a block of N pixels in one pulse, one hit per pixel per pulse) to the pixel-level simulation, at the detection-optimal pulse period of each case. For a spot of 0.6 to 1.2 pixel (rms) the simulated detection probability per pulse is close to the analytical one (0.86--0.93 against 0.84 in the near case, 0.49--0.61 against 0.44 in the median case and 0.06--0.11 against 0.058 in the far case), and the fraction of the signal in the best block is 0.60--0.95, against the assumed C = 0.7. For a more concentrated spot the analytical model is optimistic: a pixel records one hit per pulse, so m = 5 hits require the signal to reach at least five pixels. For the assumed PSF ($\sigma_{PSF}$ = 1.3 mrad, 0.45 pixel) the simulated probability is 0.54, 0.23 and 0.034 in the near, median and far cases, and the time to reach a 95\% detection probability is 23 \textmu{}s, 0.45 ms and 3.5 ms instead of 10 \textmu{}s, 0.20 ms and 2.0 ms for the detection alone (the 12 \textmu{}s of Section 5.1 also includes the time to collect the counts for the angle). The detection times of Section 5.1 are therefore optimistic by up to a factor of about 2.3 for a spot of 0.45 pixel. I made this check for the Earth case only; the lunar and Earth-Moon times come from the same analytical model and are expected to be optimistic in the same way.

The three estimators are centre-of-mass (centroid) estimators of the spot centre $\hat{x}$ from the pixel positions $x_{i}$:

\begin{gather}
\hat{x}_{B1} = \frac{\sum_i f_i\, x_i}{\sum_i f_i} \notag\\
\hat{x}_{B2} = \frac{\sum_i w_i\, x_i}{\sum_i w_i} \notag\\
w_i = -\ln(1-f_i) \notag\\
\hat{x}_{L} = \frac{\sum_i q_i\, x_i}{\sum_i q_i}
\end{gather}

\begin{equation}
\sigma_{AoA} \simeq \frac{\sigma_{PSF}}{\sqrt{N_\gamma}}
\end{equation}

where $f_{i}$ is the fraction of pulses in which pixel i fired, $q_{i}$ is the calibrated charge estimate obtained from the ToT measurement above the 5$\sigma_{e}$ threshold (linear mode), and $N_{\gamma}$ the number of detected counts in the cluster. Assuming Poisson photon arrivals, independent pixels and at most one hit per pixel per pulse, a binary pixel fires with probability f = 1 $-$ $e^{-\lambda}$, so $-$ln(1$-$f) = $\lambda$ is the true mean count (inversion of the Poisson zero-count probability, the pile-up correction used in photon counting). B2 therefore corrects the saturation (pile-up) bias of B1 near the spot centre, where f approaches 1; in the simulation f is capped at (N$_{p}$ $-$ 0.5)/N$_{p}$ for N$_{p}$ = 20 pulses, so that the inversion stays finite when a pixel fires in every pulse. The second formula is the ideal limit, the SE of the mean (the $1/\sqrt{N}$ law); the simulation adds pixel gain $g_{i}$ = 1 + $\varepsilon_{i}$, crosstalk, dead pixels and jitter, which are the scenarios plotted in Fig. 4.

These errors are obtained with 20 pulses at Tp = 40 \textmu{}s (0.8 ms of integration), which with 10 mm gives about 170 detected counts in the median case and 78 in the far case; about 75 detected counts are needed for 0.15 mrad, against the roughly 19 counts of the 0.3 mrad requirement used for Fig. 3. Without systematics the error is 0.12--0.21 mrad with the corrected binary estimator and 0.07--0.61 mrad with ToT (near, median and far). With realistic systematics (gain 3\%, crosstalk 1\%, dead pixels 0.5\%, jitter 0.1 mrad): 0.13--0.23 mrad (binary) and 0.10--0.65 mrad (ToT); in the median case 0.187 $\pm$ 0.001 and 0.356 $\pm$ 0.001 mrad. In the pessimistic case (10\%, 5\%, 2\%, 0.3 mrad): 0.19--0.28 and 0.19--0.81 mrad. Each value is the rms of 3\ensuremath{\times}10\textsuperscript{4} trials, and the SE is at most 0.001 mrad for the binary estimator and 0.003 mrad for ToT (Section 4.9), so the ranges reflect the three distances. With 10 mm the binary estimator meets the 0.3 mrad requirement in all three cases, even in the pessimistic one. The ToT estimator is better only in the near case: in the median and far cases the signal is a few counts per pixel and the noise of the linear chain (50 e\textsuperscript{$-$}, threshold at 5$\sigma$) makes it worse than the binary one, above 0.3 mrad even without systematics. With the hypothetical 30 mm aperture ToT would be better in all three cases (0.03--0.12 mrad without systematics).

\begin{figure}[htbp]
\centering
\includegraphics[width=\linewidth]{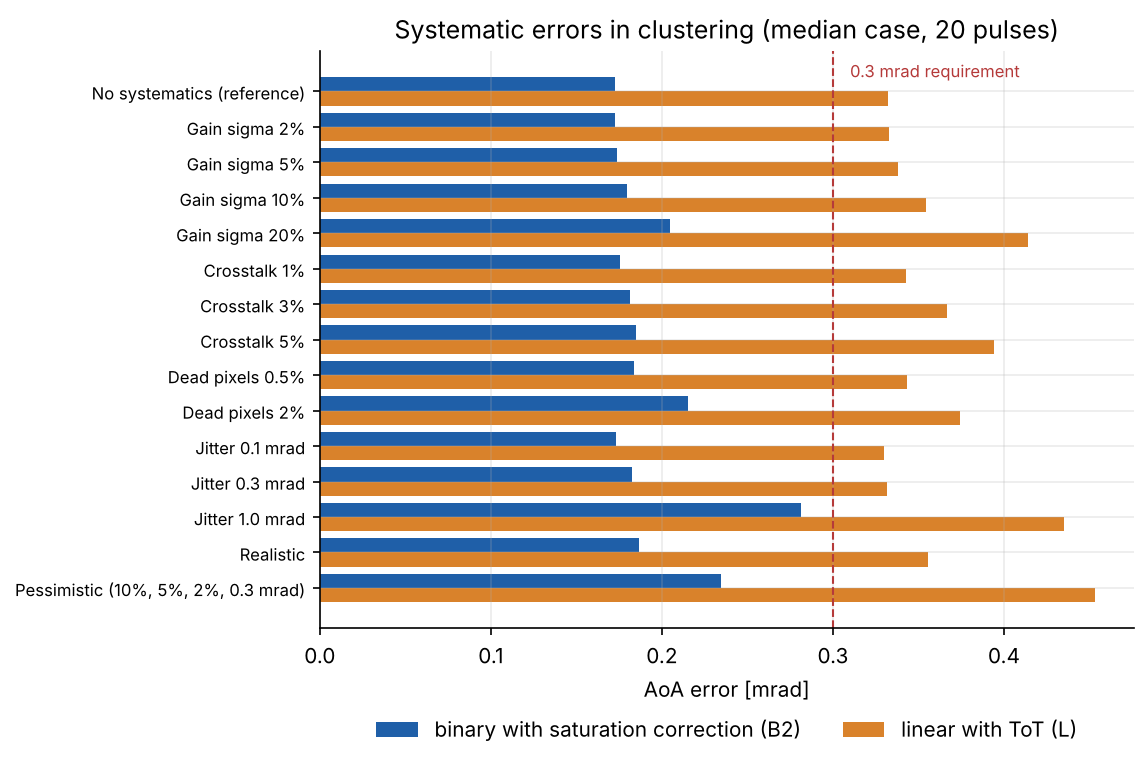}
\caption{AoA error in the median case for each systematics scenario, binary (B2) and ToT (L); dashed line: 0.3 mrad requirement. The SEs (3\ensuremath{\times}10\textsuperscript{4} trials) are at most 0.003 mrad and are smaller than the symbols.}
\end{figure}

\subsection{Jitter and per-hit attitude correction}

The beacon is fixed and the detector sees it moved by the jitter j(t), modelled as an Ornstein--Uhlenbeck (OU) process [19] (a stationary Gaussian process with exponential correlation) with standard deviation $\sigma_{j}$ and correlation time $\tau$. The attitude sensor gives a noisy measure, and each hit is corrected with the measure at its own pulse n:

\begin{gather}
\tilde{j}_n = (1+s)\, j(t_n+\delta t) + b + w_n \notag\\
x_i' = x_i - \tilde{j}_n \notag\\
\tau = \frac{1}{2\pi f_c}
\end{gather}

To first order the residual error of the corrected estimate, and the size of the jitter change over a time $\Delta$t, are:

\begin{gather}
\varepsilon_n \simeq -\left( s\, j(t_n) + \left[ j(t_n+\delta t) - j(t_n) \right] + w_n \right) \notag\\
\sigma_{\Delta}(\Delta t) = \sigma_j \sqrt{2\left(1-e^{-\Delta t/\tau}\right)} \simeq \sigma_j \sqrt{4\pi f_c\,\Delta t}
\end{gather}

Without correction the error at handover is the jitter change over the latency, $\sigma_{\Delta}$(L). With correction it is set by the synchronisation error, $\sigma_{\Delta}$($\delta$t), by the scale error, s\ensuremath{\cdot}$\sigma_{j}$, and by the white noise of the sensor; the common bias b cancels in the command. Requiring each term to stay below a budget $\sigma_{\mathrm{max}}$ gives the tolerances:

\begin{gather}
\delta t_{max} \simeq \frac{\tau}{2}\left(\frac{\sigma_{max}}{\sigma_j}\right)^{2} \notag\\
s_{max} \simeq \frac{\sigma_{max}}{\sigma_j}
\end{gather}

where:

\begin{itemize}[leftmargin=*, itemsep=2pt]
\item s is the scale error of the attitude sensor, $\delta$t the offset between the attitude clock and the hit time-stamps, b = 0.05 mrad the constant bias and $w_{n}$ = 0.05 mrad the white noise of the measure;
\item $f_{c}$ is the jitter bandwidth (100 Hz or 1 kHz), taken as the $-$3 dB cutoff frequency of the first-order low-pass spectrum of the Ornstein--Uhlenbeck process, $S(f) \propto 1/[1+(f/f_{c})^{2}]$, which gives $\tau = 1/(2\pi f_{c})$; L is the latency between the last pulse and the pointing command (1 or 10 ms);
\item $\sigma_{\mathrm{max}}$ \ensuremath{\approx} 0.3 mrad is the AoA requirement.
\end{itemize}

With $\sigma_{\mathrm{max}}$ = 0.3 mrad these formulas give about 70 \textmu{}s (1 mrad, 100 Hz), 8 \textmu{}s (3 mrad, 100 Hz) and 0.8 \textmu{}s (3 mrad, 1 kHz) for the synchronisation, and 30\% and 10\% for the scale (1 and 3 mrad).

A jitter of 0.3 mrad brings the error at handover to 0.22--0.48 mrad (binary) without correction (latency 1--10 ms; Fig. 5), and a jitter of 1--3 mrad to 0.43--4.1 mrad. Per-hit correction, with the attitude measured with 0.05 mrad of noise and bias, brings the binary error back to 0.13--0.27 mrad regardless of the jitter (0.13--0.15, 0.20--0.21 and 0.26--0.27 mrad in the near, median and far cases), below the 0.3 mrad requirement. The corrected ToT error is 0.11--0.12 mrad in the near case but 0.46--0.48 mrad in the median and 0.98--0.99 mrad in the far case, for the reason given in Section 5.2. The SEs of these values (3\ensuremath{\times}10\textsuperscript{4} trials per configuration) are at most 0.002 mrad for the corrected binary estimates, 0.005 mrad for the corrected ToT ones and 0.016 mrad (0.4\%) for the uncorrected ones.

In the median case, with the binary estimator, the correction tolerates attitude scale errors up to about 18--21\% for a jitter of 1 mrad and 5.9--6.8\% for 3 mrad, and requires synchronisation between hits and attitude to within about 38 \textmu{}s for slow jitter (1 mrad, 100 Hz), 4.3 \textmu{}s for 3 mrad at 100 Hz and 0.50 \textmu{}s for 3 mrad at 1 kHz. I use the binary estimator here because the ToT one is already above 0.3 mrad in this case (Section 5.2).

The simulated values in Fig. 6 (38, 4.3 and 0.50 \textmu{}s) are lower than the formulas (70, 8 and 0.8 \textmu{}s) because the other error terms (noise, estimator, systematics) already use part of the budget: the error without synchronisation or scale errors is 0.20 mrad, so about 0.22 mrad of the 0.3 mrad remains in quadrature. These thresholds are where the error crosses 0.3 mrad; their statistical uncertainty, from a bootstrap over the batches of trials, is about 1\% (68\% interval 37.3--38.1 \textmu{}s for the first value).

\begin{figure}[htbp]
\centering
\includegraphics[width=\linewidth]{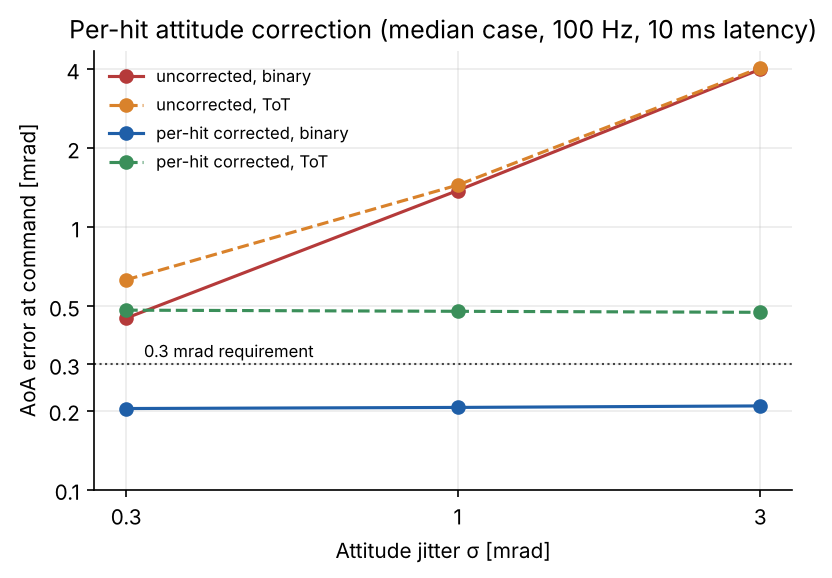}
\caption{AoA error at handover versus attitude jitter, with and without per-hit correction (median case, 100 Hz band, 10 ms latency); the SEs, at most 0.016 mrad, are smaller than the symbols.}
\end{figure}

\begin{figure}[htbp]
\centering
\includegraphics[width=\linewidth]{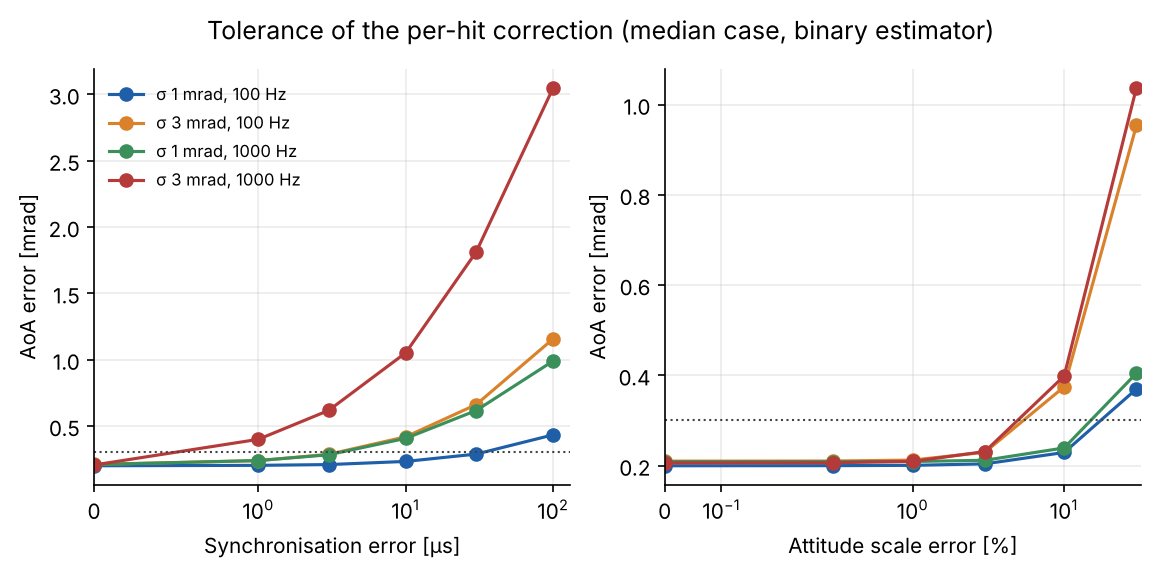}
\caption{tolerance to attitude synchronisation errors (left) and scale errors (right), binary estimator, median case.}
\end{figure}

\textbf{Note.} The jitter model used here is an ideal model of correlated jitter: a stationary Gaussian Ornstein--Uhlenbeck process with a single bandwidth, an attitude sensor with white noise and constant bias, and a clock shared between the attitude sensor and the hit time-stamps. Its validation with real jitter data is left to future work.

\subsection{Radiation and cooling}

Proton displacement damage raises the dark count linearly with the fluence $\Phi$ (a linear damage-factor relation of the kind introduced in [18] for lifetime damage) and cooling lowers it exponentially (an empirical halving rule, of Arrhenius type):

\begin{gather}
\mathrm{DCR}(\Phi,T) = \mathrm{DCR}_0 + \frac{P_{trig}}{q}\,\frac{K\,\Phi}{c(T)} \notag\\
c(T) = 2^{(296\,\mathrm{K}-T)/T_{1/2}} \notag\\
T_{1/2} = 8\ \mathrm{K}
\end{gather}

where K is the damage-induced dark current per pixel and per unit fluence (in A per p/cm\textsuperscript{2} per pixel), obtained from the measurement on InGaAs PIN photodiodes scaled with the pixel area and treated as an ideal constant: linear in the fluence, independent of the proton energy beyond the 100 MeV equivalent, of the bias and of annealing; $P_{\mathrm{trig}}$ the avalanche triggering probability (assumption), q the electron charge, $\mathrm{DCR}_{0}$ = 10\textsuperscript{4}/s the pre-irradiation value, and c(T) the cooling factor with respect to 296 K. The damage-induced term of each pixel is multiplied by a lognormal factor with $\sigma$ = 1.5, a width assumed here (the Appendix gives the effect of $\sigma$ = 1.0 and 2.0 on the life); the factor is normalised to unit mean, so that the spread redistributes the damage among the pixels without changing its mean. A pixel is masked when DCR > 20\ensuremath{\cdot}$\mathrm{DCR}_{0}$, and the end of life is reached when more than 10\% of the pixels are masked or the mean DCR of the kept pixels exceeds 3\ensuremath{\cdot}$\mathrm{DCR}_{0}$, with $\Phi$ = $\Phi_{\mathrm{yr}}$ \ensuremath{\times} (years). This produces the curves of Fig. 7.

For Fig. 8 the detector temperature $T_{d}$ follows from the heat lifted by the TEC and the heat rejected by the radiator:

\begin{gather}
Q_c = P_{det} + G_{par}\,(T_{bus}-T_d) + \epsilon_{eff}\,\sigma_{SB}\,A_{det}\left(T_{dome}^{4}-T_d^{4}\right) \notag\\
Q_h = Q_c + P_{TEC} \notag\\
P_{TEC} \simeq \frac{Q_c\,(T_r-T_d)}{\eta_C\,T_d}
\end{gather}

\begin{gather}
C_r\,\frac{dT_r}{dt} = Q_h + A_r\left(\alpha_r\,S\cos\zeta\,(1-\kappa) + q_{par}\right) - \epsilon_r\,\sigma_{SB}\,A_r\,T_r^{4} \notag\\
1/c(T_{eff}) = \left\langle \frac{1}{c(T_d)} \right\rangle \ \Rightarrow\ T_{eff} = 296\,\mathrm{K} + T_{1/2}\,\log_2\left\langle \frac{1}{c(T_d)} \right\rangle
\end{gather}

where $Q_{c}$ is the heat load on the detector ($P_{\mathrm{det}}$ its dissipation, $G_{\mathrm{par}}$ the parasitic conductance to the bus at $T_{\mathrm{bus}}$, the last term the radiative load from the dome at $T_{\mathrm{dome}}$, following the Stefan--Boltzmann law), $Q_{h}$ the heat to reject, $P_{\mathrm{TEC}}$ the electrical power of the TEC with Carnot fraction $\eta_{C}$ (the fraction of the ideal Carnot coefficient of performance that is achieved), $C_{r}$ the thermal capacity of the radiator plate (a lumped-capacitance heat balance), $A_{r}$ its area, $\alpha_{r}$ and $\varepsilon_{r}$ its solar absorptivity and emissivity, S the solar flux, $\zeta$ the sun angle, $\kappa$ the fraction shaded by the sun shield, $q_{\mathrm{par}}$ the parasitic flux, and $\sigma_{\mathrm{SB}}$ the Stefan-Boltzmann constant. $T_{\mathrm{eff}}$ is the constant temperature that would give the same mean dark count as the real, varying $T_{d}$. Since the dark count is proportional to $1/c(T)$, it is the inverse of the cooling factor that must be averaged, not $c$ itself (averaging $c$ would underestimate the dark count when $T_{d}$ varies). $T_{\mathrm{eff}}$ is the quantity plotted in Fig. 8 at the worst beta angle, and the lifetime of Fig. 7 is read at that temperature.

At room temperature the detector degrades within a few months at LEO fluences. I took the damage-equivalent fluences from the non-ionizing energy loss tool of the Space Environment Information System (SPENVIS) web tool, run on the orbit-averaged proton spectra of the AP9 model [30] (IRENE, version 1.50), for a circular orbit at 550 km and 53\textdegree{}, a one-year mission, in the mean and 95th-percentile modes. The tool passes the spectra through spherical aluminium shielding and weights them with a silicon displacement-damage curve (JPL, 1996) relative to 100 MeV protons. The fluences of 100 MeV-equivalent protons are 3.0\ensuremath{\times}10\textsuperscript{9}, 1.4\ensuremath{\times}10\textsuperscript{9} and 6.9\ensuremath{\times}10\textsuperscript{8} p/cm\textsuperscript{2} per year (mean) and 6.3\ensuremath{\times}10\textsuperscript{9}, 2.8\ensuremath{\times}10\textsuperscript{9} and 1.4\ensuremath{\times}10\textsuperscript{9} (95th percentile) for 0.25, 1 and 4 g/cm\textsuperscript{2}. I checked them with my own calculation, with continuous slowing down in aluminium and a damage function anchored to silicon data, which is also the one that I use for the cislunar case in Section 6.2: it agrees within 10\% at 0.25 g/cm\textsuperscript{2} and within 4\% at 1 and 4 g/cm\textsuperscript{2}, and repeating it with different interpolations of the spectra changes it by 10--15\% more. A first calculation with the AP8 model [17], with my own orbit code and the IGRF field, gave 2.5\ensuremath{\times}10\textsuperscript{9}, 1.5\ensuremath{\times}10\textsuperscript{9} and 6.3\ensuremath{\times}10\textsuperscript{8} for the same shieldings (solar minimum for 0.25 and 1 g/cm\textsuperscript{2}, maximum for 4), about 20\% from the AP9 mean. With cooling to 233 K the service life is 4.0, 8.6 and 17 years for the mean and 1.9, 4.3 and 8.8 years for the 95th percentile, for 0.25, 1 and 4 g/cm\textsuperscript{2}. A 5-year life at 233 K therefore needs about 0.4 g/cm\textsuperscript{2} of aluminium-equivalent shielding or more in the mean environment and about 1.3 g/cm\textsuperscript{2} at the 95th percentile. These lifetimes depend strongly on the assumptions: for a reference annual fluence of 1.5\ensuremath{\times}10\textsuperscript{9}, a dark-count halving every 6 K or every 10 K, instead of 8 K, gives about 49 or 2.7 years. The fluences are those of a public model run by a web tool, not my own result: if the model is wrong, so are these values. The results leave out solar particle events. The fluences come from AP9 and the SPENVIS non-ionizing energy loss tool, which I used without modifying them; the damage per proton, the fraction of masked pixels and the halving of the dark-count rate every 8 K are my own models or assumptions, calibrated on an InGaAs PIN photodiode and not on a SPAD array. The service lives therefore inherit the uncertainty of both the environment models (AP9 does not include solar particle events) and the damage models.

The thermal budget shows that 233 K is reachable with a TEC and a radiator of at least 0.1 m\textsuperscript{2} with a sun shield of at least 50\% and a dissipation of up to 5 W; with 0.05 m\textsuperscript{2} the detector rises to about 248 K. A passive conductive link of 0.5 W/K without a TEC gives 255, 234 and 219 K for 0.05, 0.1 and 0.2 m\textsuperscript{2} in the same conditions (Fig. 8), so in this model a radiator of 0.1--0.2 m\textsuperscript{2} may make the TEC unnecessary. The capacity of the TEC is the critical parameter: with 6 W instead of 12 W the detector is not held at 233 K (Table 6).

\begin{figure}[htbp]
\centering
\includegraphics[width=\linewidth]{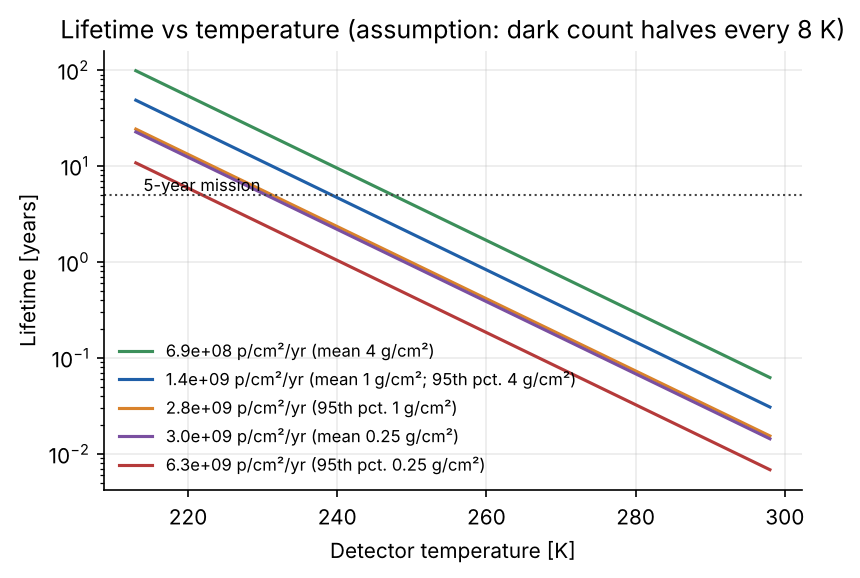}
\caption{detector service life versus temperature for annual fluences from AP9 (550 km, 53\textdegree{}; aluminium shielding of 4, 1 and 0.25 g/cm\textsuperscript{2}; the lines are labelled by annual fluence, with the mean and 95th-percentile cases that give it). Criterion: more than 10\% of pixels masked or mean DCR above 3 times nominal.}
\end{figure}

\begin{figure}[htbp]
\centering
\includegraphics[width=\linewidth]{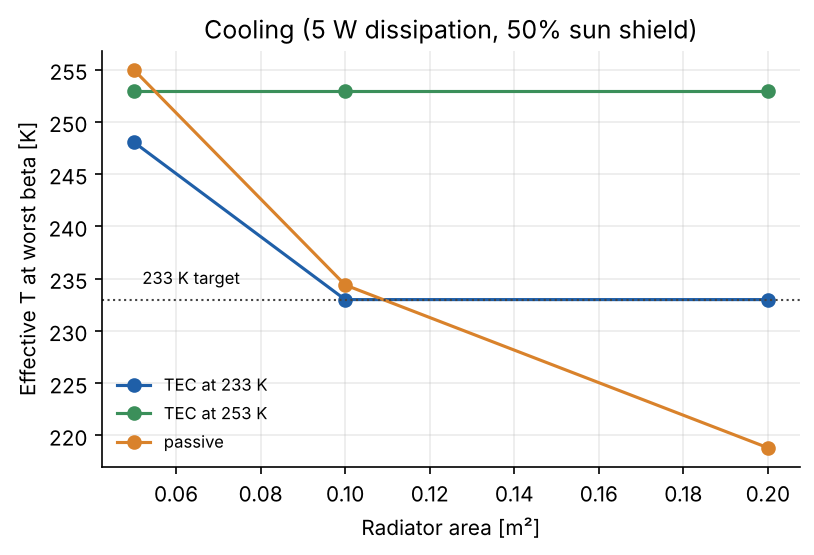}
\caption{effective detector temperature at worst-case beta versus radiator area (5 W, 50\% sun shield). The passive curve is a conductive link of 0.5 W/K to the radiator, without a TEC.}
\end{figure}

\subsection{Background: Sun, Earth, Moon, stars}

\begin{itemize}[leftmargin=*, itemsep=2pt]
\item \textbf{Sun:} about 2\ensuremath{\times}10\textsuperscript{13} counts/s with 10 mm and a 1 nm filter (2\ensuremath{\times}10\textsuperscript{14} with 30 mm), concentrated on about 8 pixels; in the useful field for 62--100\% of the orbit, so the mask is always active. With Harvey-type stray light the mask is 0.3--0.8\textdegree{} (B1 from 10\textsuperscript{$-$5} to 10\textsuperscript{$-$3} sr\textsuperscript{$-$1}; the lower value is a floor set by the solar disc and margin), but with a tighter tolerance (10\textsuperscript{4}/s) it would grow to 0.5--4.5\textdegree{}.
\item \textbf{Earth near the limb:} from 1.9\ensuremath{\times}10\textsuperscript{2} to 3.2\ensuremath{\times}10\textsuperscript{4} counts/s per pixel at 1\textdegree{} with 10 mm, always below the tolerance of 3.2\ensuremath{\times}10\textsuperscript{5} (32 times the nominal dark count and below the pixel saturation of 10\textsuperscript{6}/s) by a factor of 10 at the worst corner (B1 = 10\textsuperscript{$-$3}, reflectance 0.5). The stray-light model assumes a smaller solid angle for the Earth disc than its real one, so these values may be up to about three times higher; the worst corner would still be below the tolerance. With the hypothetical 30 mm they would be nine times higher, up to 2.9\ensuremath{\times}10\textsuperscript{5} against a tolerance of 4.6\ensuremath{\times}10\textsuperscript{5}, with little margin.
\item \textbf{Stars and planets:} up to about 2\ensuremath{\times}10\textsuperscript{3} counts/s per pixel (Venus) with 10 mm, below the dark counts and maskable with ephemerides.
\item \textbf{Zodiacal light:} about 0.2 counts/s per pixel, negligible.
\end{itemize}

\subsection{Sensitivity to aperture and correlated noise}

Two assumptions of the nominal results are tested here: the 10 mm aperture, which is slightly above what the optical constraints of Section 3.2 allow for a single lens per pixel group, and the independence of the dark counts. For the aperture I repeat the optimisation with other values. The acquisition time includes the time needed to collect the counts for the AoA estimate: with the 0.3 mrad target used in the paper, 5 mrad beacon and DCR 10\textsuperscript{4}/s, the near, median and far cases take about 12 \textmu{}s, 0.2 ms and 2.0 ms with 10 mm, against 1.3, 14 and 31 \textmu{}s with the hypothetical 30 mm. With the apertures that the optics allow (6--9 mm, Section 3.2), the median and far times are 0.36--5.1 ms and 4.3--116 ms. If the AoA target is tightened to 0.15 mrad (about 75 counts), the times are 46 \textmu{}s, 0.5 ms and 2.0 ms with 10 mm, and 5, 57 and 123 \textmu{}s with 30 mm.

For the noise I add optical crosstalk: each primary dark count fires each neighbouring pixel inside the same block with probability $p_{\mathrm{xt}}$ within the coincidence window. I take the worst-case block position (3 neighbours in a 2\ensuremath{\times}2 block, 8 in a 3\ensuremath{\times}3 block), and I re-optimise the architecture, m and W so that the whole-frame FAR stays below 0.01/s. Afterpulsing arrives microseconds after the primary hit, outside the window and during the dead time, so it only raises the effective DCR by the factor (1 + $p_{\mathrm{ap}}$); for $p_{\mathrm{ap}}$ = 10\% it changes the results by at most about 2\%. Crosstalk matters much more. With 10 mm, $p_{\mathrm{xt}}$ = 1\% and 3\% increase the near, median and far times to 18 \textmu{}s, 8 ms and 0.78 s for 1\%, and to 29 \textmu{}s, 92 ms and 21 s for 3\%. With the hypothetical 30 mm the same values give about 2, 22 and 55 \textmu{}s and 3, 36 and 154 \textmu{}s. At $p_{\mathrm{xt}}$ = 10\% no combination satisfies the false-alarm limit. The reason is that crosstalk makes a single dark count look like a small coincidence, so m has to rise to 8--9 hits and the beacon must be strong enough to deliver them; this penalises the weak-signal (far, small-aperture) cases most.

From these simulations I conclude that:

\begin{enumerate}[leftmargin=*, itemsep=2pt]
\item with the nominal 10 mm aperture the detection times are milliseconds or less, but only if crosstalk is kept at or below about 1\%; at the percent level the far case is already hundreds of milliseconds, and with 6 mm instead of 10 mm the far case is 116 ms even without crosstalk;
\item crosstalk becomes a requirement on the detector (trenches, optical isolation between pixels).
\end{enumerate}

\textbf{Note.} The crosstalk model is a worst-case model (worst-case block position, simultaneous crosstalk in all neighbours).

\section{Secondary case: lunar constellation}

\subsection{Detection and AoA performance}

Walker 65\textdegree{}:24/4/1 constellation around the Moon, 5 mrad beacon, 10 mm aperture, 0.5 W, DCR 10\textsuperscript{4}/s. The useful field is larger than in LEO (lunar half-angle 39.7\textdegree{}, with a 10 km margin above the surface, at 1000 km and 15\textdegree{} at 5000 km), so the pixel subtends 3.33 and 3.51 mrad at 1 Mpixel. As in the Earth case the spot is 0.445 pixel, i.e. about 1.5 mrad (1.6 mrad at 5000 km), so about 25 counts (28 at 5000 km) are needed for 0.3 mrad instead of 19; this sets the near-case times at 1000 km.

\begin{table}[htbp]
\centering
\footnotesize
\caption{Lunar constellation results (5\,mrad beacon, 10\,mm aperture, 0.5\,W, DCR $10^{4}$/s). The AoA errors are the rms of $3\times10^{4}$ trials $\pm$ one SE. The last block repeats the 1000 km and 5000 km cases with the hypothetical 30\,mm aperture; at 1000 km the optimiser lengthens the pulse period with distance (1, 5 and 10 \textmu{}s for the near, median and far cases), which keeps the expected counts per pulse, and so the probability per pulse, at the same value of 0.73 in all three.}
\begin{tabularx}{\linewidth}{>{\raggedright\arraybackslash}p{2.2cm} >{\raggedright\arraybackslash}X >{\raggedright\arraybackslash}X >{\raggedright\arraybackslash}X >{\raggedright\arraybackslash}X >{\raggedright\arraybackslash}X}
\toprule
\textbf{Altitude} & \textbf{Case} & \textbf{Distance [km]} & \textbf{Detection} & \textbf{P per pulse} & \textbf{AoA error binary / ToT [mrad]} \\
\midrule
1000 km & Near & 1343 & 27 \textmu{}s & 0.95 & 0.211 $\pm$ 0.001 / 0.241 $\pm$ 0.001 \\
1000 km & Median & 2972 & 0.14 ms & 0.53 & 0.213 $\pm$ 0.001 / 0.380 $\pm$ 0.002 \\
1000 km & Far & 4207 & 0.73 ms & 0.15 & 0.237 $\pm$ 0.001 / 0.572 $\pm$ 0.002 \\
5000 km & Near & 3488 & 0.25 ms & 0.38 & 0.228 $\pm$ 0.001 / 0.449 $\pm$ 0.002 \\
5000 km & Median & 9473 & 0.43 s & $3\times10^{-4}$ & 0.467 $\pm$ 0.002 / 1.33 $\pm$ 0.01 \\
5000 km & Far & 12859 & 7.3 s & $2\times10^{-5}$ & 0.672 $\pm$ 0.003 / 1.52 $\pm$ 0.05 \\
\midrule
\multicolumn{6}{l}{\textit{Hypothetical 30 mm aperture}} \\
1000 km & Near & 1343 & 3 \textmu{}s & 0.73 & 0.223 $\pm$ 0.001 / 0.319 $\pm$ 0.001 \\
1000 km & Median & 2972 & 15 \textmu{}s & 0.73 & 0.223 $\pm$ 0.001 / 0.320 $\pm$ 0.001 \\
1000 km & Far & 4207 & 30 \textmu{}s & 0.73 & 0.223 $\pm$ 0.001 / 0.322 $\pm$ 0.002 \\
5000 km & Near & 3488 & 23 \textmu{}s & 0.58 & 0.228 $\pm$ 0.001 / 0.384 $\pm$ 0.001 \\
5000 km & Median & 9473 & 0.18 ms & 0.46 & 0.226 $\pm$ 0.001 / 0.418 $\pm$ 0.001 \\
5000 km & Far & 12859 & 0.82 ms & 0.14 & 0.251 $\pm$ 0.001 / 0.621 $\pm$ 0.002 \\
\bottomrule
\end{tabularx}
\end{table}

With 10 mm the architecture works around the Moon at 1000 km altitude, with detection from 27 \textmu{}s to 0.7 ms and a binary AoA error of 0.21--0.24 mrad, below 0.3 mrad. At 5000 km altitude it works only for the nearest satellites: the median and far cases (9473 and 12,859 km) need 0.43 s and 7.3 s, with a detection probability per pulse of $3\times10^{-4}$ and $2\times10^{-5}$, and the binary AoA error rises to 0.47 and 0.67 mrad. With the hypothetical 30 mm aperture all these cases would take less than 1 ms. With the 6 and 9 mm that the optics allow (Section 3.2), at 1000 km the detection takes 75 and 34 \textmu{}s in the near case, 2.1 and 0.20 ms in the median and 31 and 1.4 ms in the far case; at 5000 km the nearest satellites are still acquired in 6.9 and 0.45 ms, but the median and far cases need 52 s and 990 s with 6 mm and 1.1 s and 20 s with 9 mm. The AoA errors of Table 3 use 20 pulses at the detection-optimal pulse period, so they rest on fewer counts per pulse than the Earth values of Section 5.2 (which use Tp = 40 \textmu{}s); they cannot be compared directly, and the larger pixel is a secondary cause of the difference. The ToT estimator is worse than the binary one in every case of this table, with both apertures.

\subsection{Thermal, radiation and background}

\textbf{Thermal.} At 1000 km lunar altitude the dome ranges from about 33 K (night, one face; idealised, with no conduction from the bus) to 203 K (noon) at beta 0, against 141--208 K around the Earth. The view factor towards the surface is 0.013 (Earth: 0.094). The detector is easier to cool; dome, filter and supports must survive the lunar night.

\textbf{Radiation.} The belts do not reach lunar orbit, so the AP9 results of Section 5.4 hold only for the Earth case; the models and assumptions that the service lives inherit are those listed there. As a first-order estimate with public models, I computed the proton fluence equivalent to 100 MeV, in the convention of Section 5.4 (aluminium shielding of constant thickness, continuous slowing down, no secondaries), in open space at 1 AU. The Moon blocks a fraction of the sky $(1-\cos\rho)/2$, 11\% at 1000 km and 1.7\% at 5000 km, which would lower the values below slightly; I give the open-space values, which are conservative. \emph{Galactic cosmic rays:} the local interstellar spectrum of protons given in [26], with force-field modulation (potential 400 MV at solar minimum and 1100 MV at maximum), protons only, multiplied by 1.5 for helium and heavier ions (assumption). The integral flux above 100 MeV is 4.7 cm\textsuperscript{-2}s\textsuperscript{-1} at solar minimum and 1.5 at maximum, of the order of the few cm\textsuperscript{-2}s\textsuperscript{-1} usually quoted for galactic cosmic rays. The annual fluence equivalent to 100 MeV is 1.3\ensuremath{\times}10\textsuperscript{8} p/cm\textsuperscript{2} at minimum and 3.6\ensuremath{\times}10\textsuperscript{7} at maximum, independent of the shielding between 0.25 and 4 g/cm\textsuperscript{2}, and 5 to 23 times below the AP9 mean values of the Earth case (cosmic rays at solar minimum); alone, it would give a life of about 90 years at 233 K. \emph{Solar energetic particles} dominate. For them I used the ESP-PSYCHIC model [27], with the spectra that I generated by running it in the SPENVIS web tool (version 4.6) for a 2-year mission (1.87 years at solar maximum), outside the magnetosphere and without magnetic shielding, at 90, 95 and 99\% confidence that the fluence is not exceeded; I took the integral proton spectrum from 1 to 500 MeV and passed it through the shielding and damage calculation of Section 5.4. The total fluences equivalent to 100 MeV for 2 years are 8.3\ensuremath{\times}10\textsuperscript{10} (90\%), 1.4\ensuremath{\times}10\textsuperscript{11} (95\%) and 3.4\ensuremath{\times}10\textsuperscript{11} (99\%) p/cm\textsuperscript{2} behind 0.25 g/cm\textsuperscript{2} of aluminium; 1.9\ensuremath{\times}10\textsuperscript{10}, 3.4\ensuremath{\times}10\textsuperscript{10} and 1.0\ensuremath{\times}10\textsuperscript{11} behind 1 g/cm\textsuperscript{2}; and 3.2\ensuremath{\times}10\textsuperscript{9}, 6.4\ensuremath{\times}10\textsuperscript{9} and 2.3\ensuremath{\times}10\textsuperscript{10} behind 4 g/cm\textsuperscript{2}. The end-of-life damage budget of the model of Section 5.4 at 233 K is 1.2\ensuremath{\times}10\textsuperscript{10} p/cm\textsuperscript{2} (8.0 years at a reference 1.5\ensuremath{\times}10\textsuperscript{9} p/cm\textsuperscript{2} per year). With 0.25 g/cm\textsuperscript{2} the solar events alone use 7 to 28 times the budget, so the detector does not survive; with 1 g/cm\textsuperscript{2} they use 1.5 times (90\%) to 8.7 times (99\%) the budget; with 4 g/cm\textsuperscript{2} they use 27\% (90\%), 54\% (95\%) and 1.9 times (99\%) the budget (Figure 9). The shielding at which the solar events alone use up the budget is about 1.4, 2.5 and 6.8 g/cm\textsuperscript{2} at 90, 95 and 99\%, and the shielding at which they use half of the budget is about 2.5, 4.2 and 11 g/cm\textsuperscript{2}. With 4 g/cm\textsuperscript{2} the life with cosmic rays at solar minimum is about 69 years after a 90\% event and zero after a 99\% one. In this estimate the life in lunar orbit is therefore set by solar events: about 4 g/cm\textsuperscript{2} of aluminium-equivalent shielding is needed to keep a margin against the 90--95\% event, and the 99\% event would need about 7 g/cm\textsuperscript{2} or more. The numerical integration over energy uses a fine grid, because the fluence behind the shield depends on the steep part of the solar spectrum. The estimate uses the ESP-PSYCHIC spectrum for a 2-year mission, valid up to 500 MeV; the displacement-damage function of Section 5.4, which is that of silicon, with a plateau at high energy; and no secondaries in the shield, which become important at the larger thicknesses. It omits the lunar albedo and the passage through the magnetotail. The same solar events also reach the Earth case, where they are not included (Section 5.4) and where the geomagnetic field screens them only in part, so the comparison with the AP9 values is only indicative.

\begin{figure}[htbp]
\centering
\includegraphics[width=\linewidth]{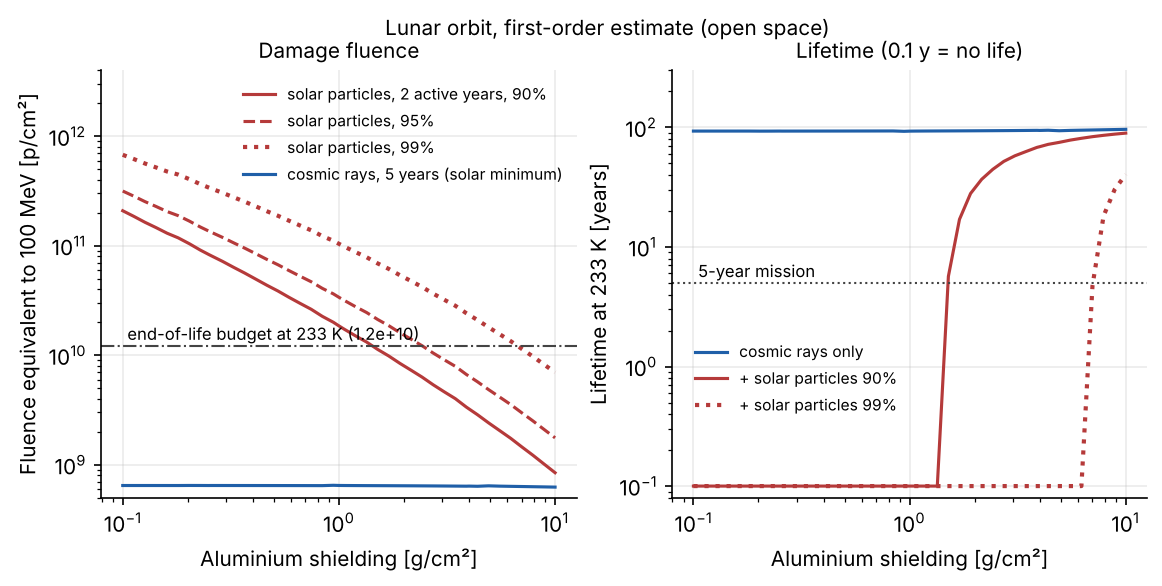}
\caption{First-order radiation estimate for a lunar orbit, in open space (conservative, no sky blocked by the Moon). Left: fluence equivalent to 100 MeV as a function of the aluminium shielding, for the solar events of a 2-year mission (ESP-PSYCHIC, SPENVIS) at 90, 95 and 99\% confidence and for 5 years of cosmic rays at solar minimum; the dash-dotted line is the end-of-life damage budget at 233 K. Right: resulting life at 233 K (0.1 years means that the detector does not survive the event). Assumptions: ESP-PSYCHIC spectra and cosmic-ray spectrum as in the text, factor 1.5 for heavy ions, constant-thickness aluminium, no secondaries.}
\end{figure}

\textbf{Background.} The Moon subtends the 39.7\textdegree{} cone (1000 km) excluded from the field; the Sun and Earth remain as sources to be masked. The Earth disc seen from lunar orbit is treated in Section 7.2.

\subsection{Why the lunar case suits the architecture}

I see four reasons, all of which follow from the results above. First, the geometry is favourable: a lunar constellation has few satellites and distances of a few thousand km, so the same 0.5 W, 10 mm and 5 mrad beacon used on Earth is detected in 27 \textmu{}s to 0.7 ms at 1000 km altitude (up to 7 s at 5000 km), without a larger transmitter. Second, the thermal problem is easier: the view factor towards the surface is 0.013 against 0.094 for the Earth, which helps the detector reach the temperature it needs with a smaller radiator. Third, there is no practical alternative on the ground side: with the satellites far from continuous ground contact, a terminal that finds its partner autonomously and directly measures the direction is more valuable than one that depends on a scan or on commanded pointing. Fourth, the lunar infrastructure is still being defined, so an acquisition concept can be proposed before the terminals are frozen. These are arguments for the scenario, not proof: the cislunar radiation is only a first-order estimate (Section 6.2) and the lunar-night survival of the optics is not modelled here.

\section{Additional case: Earth-Moon link}

Distance between centres 356,500--406,700 km (median 384,400 km); the near and far cases subtract and add the orbital radii of the two satellites (346,842 and 416,358 km, Section 4.1). The receiver sits on the Earth satellite; the beacon comes from the lunar satellite.

\subsection{Link budget and beacon divergence}

Fig. 10 uses the same chain as Section 5.1 ($r_{s}$, $\mu$, $P_{d}$, $t_{\mathrm{acq}}$) with R equal to the Earth-Moon distance of 384,400 km and a beacon divergence $\theta$ of 100 \textmu{}rad to 5 mrad. Because $r_{s}$ \ensuremath{\propto} $P_{\mathrm{tx}}$ d\textsuperscript{2}/($\theta$R)\textsuperscript{2}, reducing $\theta$ by a factor 2 raises the counts per pulse by a factor 4, and this is why the detection time falls by orders of magnitude between 5 mrad and 200 \textmu{}rad. The transmit pointing requirement and the point-ahead angle that go with the divergence are given below. Because the receiver moves during the light time $\tau = R/c$, the transmitter must point ahead of the line of sight it sees. I call $\theta_{pa}$ the geometric angle between the instantaneous line of sight and the direction in which the transmitter must point; the angle that the receiver observes between the direction of arrival of the light and the direction in which it must transmit back is about twice as large, $\theta_{pa,obs} \simeq 2\,\theta_{pa}$, because the receiver has also moved and the light that arrives left the transmitter at an earlier position. This is the usual convention, in which the separation between the received and the transmitted direction is $2v_{\perp}/c$ [32]; the geometric angle is half of it. In vector form, with $\mathbf{r}$ and $\mathbf{v}$ the position and velocity vectors in an inertial frame (including the orbital motion of the Moon), the pointing direction and, to first order, the angle are:

\begin{gather}
\hat{\mathbf{u}}_{pa} = \frac{\mathbf{r}_{rx}(t+\tau)-\mathbf{r}_{tx}(t)}{c\,\tau}, \qquad c\,\tau = \left|\mathbf{r}_{rx}(t+\tau)-\mathbf{r}_{tx}(t)\right| \notag\\
\theta_{pa} \simeq \frac{\left|\left(\mathbf{I}-\hat{\mathbf{u}}\hat{\mathbf{u}}^{T}\right)\left(\mathbf{v}_{rx}-\mathbf{v}_{tx}\right)\right|}{c} \le \frac{v_{\perp,tx}+v_{\perp,rx}}{c} \notag\\
\theta_{pa,obs} \simeq 2\,\theta_{pa} \le \frac{2\,(v_{\perp,tx}+v_{\perp,rx})}{c} \notag\\
\sigma_{pt} \simeq \frac{\theta}{4}
\end{gather}

where $\hat{\mathbf{u}}$ is the unit line-of-sight vector, $(\mathbf{I}-\hat{\mathbf{u}}\hat{\mathbf{u}}^{T})$ the projector onto the plane perpendicular to it, $v_{\perp}$ the magnitudes of the transverse velocities of the two terminals, c the speed of light and $\sigma_{\mathrm{pt}}$ the pointing error allowed at the transmitter. The values of $\theta_{pa}$ = 18 \textmu{}rad (GEO) and 33 \textmu{}rad (LEO) quoted below, and the corresponding $\theta_{pa,obs}$ = 36 and 66 \textmu{}rad, use the right-hand bounds, i.e. the transverse speeds are added in magnitude, which is the worst case (aligned velocities); the vector form gives a smaller angle for other geometries.

The choice $\sigma_{\mathrm{pt}} \simeq \theta/4$ is a design convention. It places the half-width of the beam, $\theta/2$, at two standard deviations of the pointing error per axis: each axis is then inside the beam 95\% of the time, and the radial error (Rayleigh distributed) exceeds $\theta/2$ with probability $e^{-2} \approx 14\%$. For a Gaussian beam whose 1/e\textsuperscript{2} full width is $\theta$ (standard deviation $\theta$/4), a receiver displaced by one $\sigma_{\mathrm{pt}}$ still receives $e^{-1/2} \approx 61\%$ of the on-axis intensity ($-$2.2 dB). A budget of $\theta$/8 would lower the probability of leaving the half-width to $e^{-8} \approx 3\times10^{-4}$, at the cost of doubling the pointing accuracy required. The detection model treats the spot as uniform, so every time quoted below assumes that the receiver is inside the beam.

A 5 mrad beacon (spot of about 1900 km) is unusable: with 10 mm or 30 mm it is never feasible, and with 100 mm it is infeasible in most cases within the 20 \textmu{}J pulse-energy limit (the only feasible case I found is the nearest distance at 2 W, about 7.7 hours or 2.8\ensuremath{\times}10\textsuperscript{4} s, at the edge of feasibility; false alarms below 0.01/s). Detection becomes practical with beacons of the order of 1 mrad and a large aperture, and is fast with 200--500 \textmu{}rad. With 100 mm and 2 W: 18 ms at 1 mrad, 0.11 ms at 500 \textmu{}rad and 6.8 \textmu{}s at 200 \textmu{}rad. With 30 mm and 2 W: 1.9 s at 500 \textmu{}rad and 0.71 ms at 200 \textmu{}rad in the median case. With 10 mm and 2 W only the narrow beacons are feasible: 4.0, 10.6 and 23 s at 200 \textmu{}rad (near, median and far), and 7, 18 and 36 ms at 100 \textmu{}rad (0.9--5 s at 200 \textmu{}rad with 10 W). With 6 and 9 mm and 2 W the median case at 200 \textmu{}rad takes about 25 minutes and 29 s, and only 100 \textmu{}rad with 10 W brings it to 0.41 s and 9.7 ms. For powers other than 0.5 W the pulse period is searched on the same grid (80 values between 1 and 40 \textmu{}s), which does not contain the exact maximum period 20 \textmu{}J/$P_{tx}$, so these times are 4--17\% longer than the exact optimum, which is conservative.

The price is transmit pointing accuracy: a beacon of divergence $\theta$ needs a precision of about $\theta$/4, i.e. 125 \textmu{}rad at 500 \textmu{}rad and 50 \textmu{}rad at 200 \textmu{}rad. Very good ephemerides and attitude of the lunar satellite are required.

The AoA with 20 pulses and realistic systematics (binary and ToT, rms of 3\ensuremath{\times}10\textsuperscript{4} trials $\pm$ one SE) is 0.169 $\pm$ 0.001 and 0.165 $\pm$ 0.001 mrad with 200 \textmu{}rad, 2 W and 100 mm; 0.198 $\pm$ 0.001 and 0.439 $\pm$ 0.002 mrad with 500 \textmu{}rad, 2 W and 100 mm; 0.145 $\pm$ 0.001 and 0.108 $\pm$ 0.001 mrad with 100 \textmu{}rad, 10 W and 100 mm. These values assume 20 consecutive pulses at the detection-optimal pulse period of the median case (384,400 km), with the counts per pulse $\mu$ taken from the link budget above, the beacon present in every pulse and the receiver inside the beam; the spot is fixed on the detector during the 20 pulses (pixel of 2.92 mrad, spot size from the defocus of 0.4--0.8 pixel, ROI of 9\ensuremath{\times}9), the dark counts are independent (10\textsuperscript{4}/s per pixel, 10 ns window), the filter is ideal, and the systematics are those called realistic in Section 5.2 (gain 3\%, crosstalk 1\%, dead pixels 0.5\%, jitter 0.1 mrad).

The time of flight is 1.28 s. The point-ahead angle, a deterministic offset that must be applied at the transmitter, is at most about 33 \textmu{}rad (geometric; 66 \textmu{}rad observed) from LEO and 18 \textmu{}rad (36 \textmu{}rad observed) from GEO (transverse speeds added in magnitude: 7.59 km/s for a LEO transmitter or 3.07 km/s for GEO, 1.02 km/s for the orbital motion of the Moon and 1.34 km/s for a satellite in a 1000 km lunar orbit). It is small compared with the 0.3 mrad handover accuracy, but the observed value from LEO exceeds the 50 \textmu{}rad transmit-pointing budget of a 200 \textmu{}rad beacon. The offset must therefore be computed from the ephemerides and applied, and the 50 \textmu{}rad budget is kept for the residual error.

\begin{figure}[htbp]
\centering
\includegraphics[width=\linewidth]{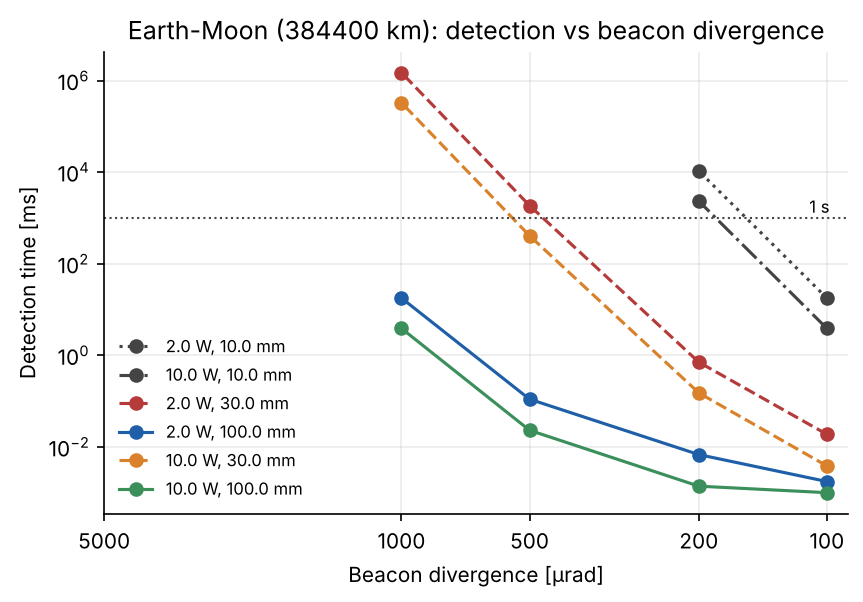}
\caption{detection time versus beacon divergence, Earth-Moon median case (384,400 km), for average power and aperture. The axis starts at 5000 \textmu{}rad, but there are no points there: at 5 mrad the median case is infeasible for every combination shown (Section 7.1), so the curves begin at 1000 \textmu{}rad.}
\end{figure}

\subsection{Geometric constraint of the Moon and Earth discs}

A disc of angular radius $\rho_{\mathrm{disc}}$ is masked with a margin of g pixels, so a satellite is lost when its projected separation from the disc centre is smaller than $s_{\mathrm{min}}$. For a circular orbit of radius a around the body, with random orientation, the fraction of time spent outside the masked disc is:

\begin{gather}
s_{min} = \left(\rho_{disc} + g\,\delta_{pix}\right) D \notag\\
f_{vis}(a) = \sqrt{1-\left(\frac{s_{min}}{a}\right)^{2}} \ \ (a \ge s_{min}) \notag\\
f_{vis}(a) = 0 \ \ (a < s_{min})
\end{gather}

where $\rho_{\mathrm{disc}}$ = $R_{\mathrm{body}}$/D is the angular radius of the disc seen by the receiver (4.5 mrad for the Moon seen from Earth, 16.6 mrad for the Earth seen from the Moon), $\delta_{\mathrm{pix}}$ the pixel angle (2.92 mrad on Earth, 3.3 mrad on the Moon), g = 1, 2 or 4 the masking margin in pixels, D the distance between the two bodies and a the orbital radius of the satellite behind the disc. A satellite with a < $s_{\mathrm{min}}$ is always inside the masked disc, which is the 0\% quoted below.

For the reverse direction, the radiance of the Earth disc seen by a lunar receiver is computed as follows. The disc is an extended Lambertian source lit by the Sun at 1550 nm (AM0 spectral irradiance $E_{\odot}$ = 0.26 W m\textsuperscript{$-$2} nm\textsuperscript{$-$1}), with SWIR reflectance $\rho$ = 0.05--0.5, so that its radiance is $L = \rho E_{\odot}/\pi$. Because the source is extended and fills the pixel, the count rate does not depend on the distance and is set by the pixel solid angle $\Omega_{\mathrm{pix}} = \delta_{\mathrm{pix}}^{2}$, the aperture area $A_{rx} = \pi d^{2}/4$, the filter bandwidth $\Delta\lambda$, and the product of optical efficiency and PDE:

\begin{gather}
r_{bg} = \eta_{opt}\,\mathrm{PDE}\;\frac{\rho\,E_{\odot}}{\pi}\;\Omega_{pix}\,A_{rx}\,\frac{\Delta\lambda}{h\nu} \notag\\
r_{bg} \le r_{tol} \ \Rightarrow\ \Delta\lambda_{max} = \Delta\lambda_{ref}\,\frac{r_{tol}}{r_{bg}(\Delta\lambda_{ref})}
\end{gather}

where $r_{\mathrm{tol}}$ = 3.2\ensuremath{\times}10\textsuperscript{5} counts/s per pixel is the largest uniform background for which the median Earth case (5 mrad beacon, 10 mm aperture, $t_{\mathrm{acq}}$ \ensuremath{\leq} 10 ms) remains feasible with a whole-frame FAR below 0.01/s. The estimate assumes full solar illumination at normal incidence and no phase function (an upper bound), an ideal rectangular filter, and no stray light.

The Moon disc (radius 4.5 mrad seen from Earth) saturates the pixels (about 4\ensuremath{\times}10\textsuperscript{6} to 8\ensuremath{\times}10\textsuperscript{6} counts/s per pixel for a Moon reflectance of 0.12, with and without a mean phase factor of 0.5; same equation as for the Earth disc below), i.e. about 12 to 24 times the tolerance (10 mm aperture, 1 nm filter), and must be masked with a margin of 1--4 pixels. A lunar satellite at 1000 km altitude (orbital radius 2737 km) never leaves the masked disc: visible 0\% of the time. The constraint relaxes for higher orbits: with a 2-pixel margin ($s_{\mathrm{min}}$ = 3982 km for D = 384,400 km), 9\% at 4000 km, 75\% at 6000 km, 92\% at 10,000 km and 98\% at 20,000 km (Fig. 11). Elliptical and near-rectilinear halo orbit (NRHO) type orbits are therefore expected to be more favourable, but the formula above is for circular orbits with random orientation and I have not simulated them. The 0\% also depends on the mask margin: it holds for a margin of about one pixel or more, and a finer pixel or per-pixel gating would change it.

In the reverse direction, a lunar receiver sees the Earth disc with a radius of 16.6 mrad and, from the equation above with $\delta_{\mathrm{pix}}$ = 3.33 mrad ($\Omega_{\mathrm{pix}}$ = 1.1\ensuremath{\times}10\textsuperscript{$-$5} sr), d = 10 mm ($A_{rx}$ = 7.9\ensuremath{\times}10\textsuperscript{$-$5} m\textsuperscript{2}), $\eta_{\mathrm{opt}}\mathrm{PDE}$ = 0.15, h$\nu$ = 1.28\ensuremath{\times}10\textsuperscript{$-$19} J and $\Delta\lambda$ = 1 nm, a radiance of 4.1\ensuremath{\times}10\textsuperscript{6} counts/s per pixel for $\rho$ = 0.05 and 4.1\ensuremath{\times}10\textsuperscript{7} for $\rho$ = 0.5, i.e. 13--130 times the tolerance (9 times more with 30 mm). Reading the signal inside the disc would therefore need a filter of 76 pm ($\rho$ = 0.05) to 7.6 pm ($\rho$ = 0.5), still unrealistic for a passive wide-field filter. A satellite in LEO is always in front of the disc or the limb: visible 0\% of the time. Medium Earth orbit (MEO; about 90--96\%) and GEO (about 96--98\%) remain acquirable.

\textbf{Consequence.} Wide-field acquisition is suitable between high orbits (NRHO, elliptical lunar frozen orbit (ELFO), MEO, GEO). Towards LEO or a low lunar orbit, ephemerides and a narrow channel are needed, with no wide-field acquisition on the disc.

\begin{figure}[htbp]
\centering
\includegraphics[width=\linewidth]{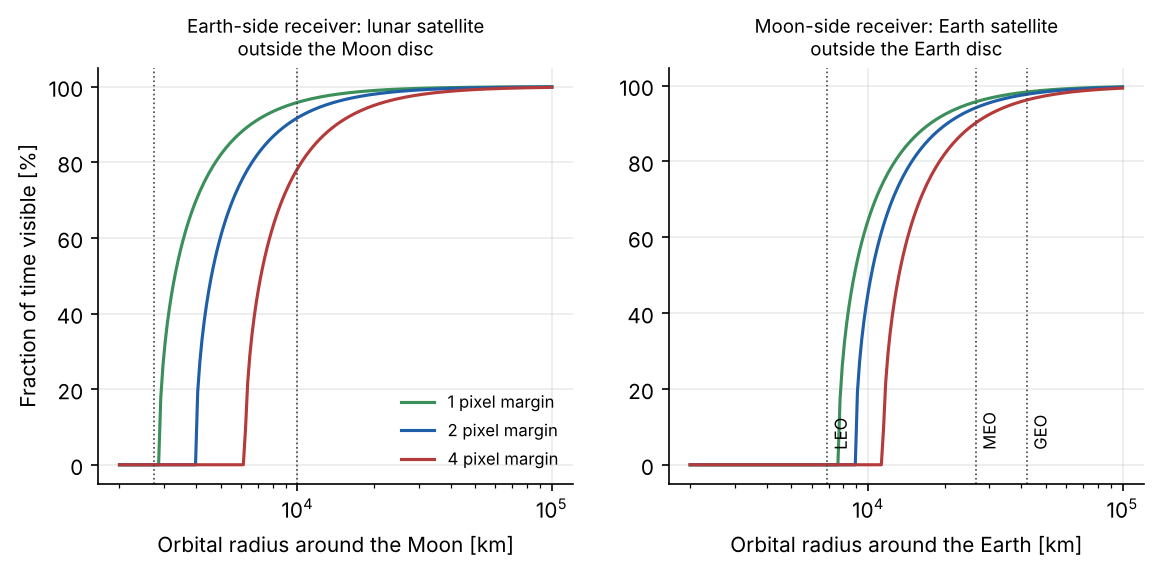}
\caption{fraction of time the satellite stays outside the masked disc, for an Earth receiver (left, lunar satellite) and a lunar receiver (right, Earth satellite), with margins of 1, 2 and 4 pixels.}
\end{figure}

\textbf{Radiation.} I did not compute a fluence specific to the Earth-Moon link. A receiver in LEO falls in the AP9 case of Section 5.4 and one in lunar orbit in the first-order estimate of Section 6.2. A receiver in GEO, which is the Earth-side terminal considered here for the point-ahead angle, is only partly shielded from solar energetic protons by the geomagnetic field: in a comparison of GOES at GEO with a spacecraft at the Sun-Earth Lagrange point 1, the median peak flux of 10--50 MeV protons at GEO is within about 20\% of the L1 value for the weaker events [31]. Its fluence is not covered by either result, and a proper estimate would need a transport calculation with secondary particles; it is outside this work (Section 8).

\subsection{Why the Earth-Moon link suits the architecture}

The one-way light time of 1.28 s makes scan-and-feedback acquisition expensive: each step of a search that waits for the partner's answer costs a round trip of about 2.6 s, and the public LCRD/ILLUMA-T results [12] (a GEO-to-LEO link, whose round trip I estimate at about 0.25 s) report a spiral acquisition scan of about 7 minutes; if the dwell of a scan is set by the round trip, each step would be about ten times slower at the Earth-Moon distance. A wide-field receiver that measures the direction from a single beacon pulse would remove the need for those steps. In the model, detection and AoA estimation are possible between high orbits with a 200--500 \textmu{}rad beacon, with detection from 7 \textmu{}s to 1.9 s and an AoA error of 0.11--0.44 mrad. The point-ahead angle (18--33 \textmu{}rad geometric, 36--66 \textmu{}rad observed) is a deterministic transmit offset that must be applied from the ephemerides; the 50 \textmu{}rad transmit-pointing budget of a 200 \textmu{}rad beacon is kept for the residual error. The disc constraint is a real limit, but it also tells where the architecture belongs: links between elliptical or NRHO-type lunar orbits and MEO or GEO relays, which are the orbits that lunar infrastructure is expected to use; GEO-to-Moon links are studied in [13]. The cost is transmit pointing of about $\theta$/4 and good ephemerides, and this is the main assumption I would test first.

\textbf{Scope of the wide field in this case.} The Earth--Moon direction changes slowly and is predictable from ephemerides, so the large FOV adds little here beyond tolerance to attitude errors; the real benefit of the architecture in this link is the single-pulse angle measurement.

\section{Comparison of the three scenarios, discussion and limits}

\begin{table}[htbp]
\centering
\footnotesize
\caption{Comparison of the three scenarios.}
\begin{tabularx}{\linewidth}{>{\raggedright\arraybackslash}X >{\raggedright\arraybackslash}X >{\raggedright\arraybackslash}X >{\raggedright\arraybackslash}X}
\toprule
 & \textbf{Earth (Walker 53\textdegree{}, 550 km)} & \textbf{Moon (Walker 65\textdegree{}, 1000--5000 km)} & \textbf{Earth-Moon} \\
\midrule
Distances & 1002--4910 km & 1343--12,859 km & 356,500--406,700 km \\
Useful beacon & 5 mrad, 10 mm, 0.5 W & 5 mrad, 10 mm, 0.5 W (1000 km altitude) & 200--500 \textmu{}rad with 100 mm and 2 W (200 \textmu{}rad with 30 mm; 100--200 \textmu{}rad with 10 mm) \\
Detection & 12 \textmu{}s--2 ms (up to about 2.3 times that for a 0.45-pixel spot, Section 5.2); 14 \textmu{}s--116 ms with 6--9 mm & 27 \textmu{}s--0.7 ms at 1000 km (34 \textmu{}s--31 ms with 6--9 mm); up to 7 s at 5000 km (up to 990 s with 6--9 mm) & 7 \textmu{}s--1.9 s with 30--100 mm, depending on divergence and aperture; with 6--9 mm only at 100 \textmu{}rad and 10 W (median case 9.7--408 ms) \\
AoA error, binary / ToT & 0.13--0.23 / 0.10--0.65 mrad & 0.21--0.24 / 0.24--0.57 mrad (1000 km) & 0.14--0.20 / 0.11--0.44 mrad (100 mm) \\
Pixel & 2.92 mrad & 3.33--3.51 mrad & 2.92 mrad (Earth side) \\
Dominant constraint & Radiation and cooling to 233 K; solar background & Night-time thermal; solar particle events set the life (first-order estimate) & Transmit pointing and the Moon and Earth discs \\
Wide-field acquisition & Possible in the model, between all satellites & Possible in the model, between lunar orbits & Only between high orbits; not towards LEO or low lunar orbits \\
\bottomrule
\end{tabularx}
\end{table}

\textbf{Limits.}

\begin{itemize}[leftmargin=*, itemsep=2pt]
\item Everything is simulation and assumption (Appendix); there is no measurement on a 1 Mpixel InGaAs detector, which does not exist in the literature today. The largest published InGaAs/InP SPAD array I found has 96\ensuremath{\times}96 pixels, 15\% PDE and a median DCR of 87 kcps at room temperature [20], so 1 Mpixel, PDE 0.30 and DCR 10\textsuperscript{4}/s at 233 K are optimistic assumptions.
\item The dome is not designed: its material, thickness, mass, contribution to the radiation shielding, anti-reflection coatings and mounting are not specified or optimised, and a dome beyond a hemisphere is difficult to manufacture and to mount (the rim must sit near the masked Earth limb).
\item The background uses simplified stray light (Harvey): ghosting, dome-filter reflections, filter drift with angle and temperature, and airglow are not modelled.
\item Radiation fluences in the Earth case come from the AP9 model (mean and 95th percentile) run in the SPENVIS web tool, including the conversion into damage, which I checked with my own simplified calculation (continuous slowing down, no secondaries); solar particle events are not included in that case; the cislunar environment is a first-order estimate with public models and SPENVIS solar-proton spectra (Section 6.2), not a transport calculation; for the Earth-Moon link there is no radiation estimate for a receiver in GEO, which lies between the two cases treated (Section 7).
\item The nominal aperture of 10 mm is slightly above what a single lens per pixel group on this detector allows (6--9 mm, Section 3.2); with 6--9 mm the median and far detection times are 0.36--5.1 ms and 4.3--116 ms. The 30 mm aperture (and, in the Earth-Moon link, the 100 mm one) that I keep as a reference would need a much larger and heavier detector or several apertures, and is far from what can be realised today.
\item The receiver does not compete with conventional terminals in sensitivity per photon: with an aperture of 6--9 mm instead of a few centimetres (for example 5 cm, my estimate) it collects roughly 30--70 times less light, about 15--18 dB. What it exchanges this for is a full-hemisphere field without scanning and a background and dark-count rate per pixel that is lower by the number of pixels. The gain therefore concerns the acquisition phase, and not data transfer or fine tracking, which remain with the narrow channel; a measurement would be needed to confirm it.
\item The comparison with published acquisition times (Section 2.4) is indicative: the references differ in uncertainty, divergence and range, my end-to-end time depends on two terms that I assume (pointing of the narrow channel and loop closure), and the 0.91 s of [36] is an equivalent time measured on the ground. The advantage also assumes a receiver on both terminals, which I have not modelled separately.
\item The attitude correction assumes OU-modelled jitter and a clock shared with the time-stamps. The detection probability assumes the beacon signal is split uniformly over the pixels of the spot, whereas the real result depends on the spot size and position inside the pixel (Section 5.2 quantifies it: up to a factor of about 2 in detection time for a spot of 0.45 pixel). The nominal FAR assumes independent dark counts; Section 5.6 tests optical crosstalk with a worst-case model that is not validated on a measured array, and hot pixels are not included in the false-alarm calculation (the radiation model of Section 5.4 masks them). Beacon identification against other sources is not treated, and the transmit pointing of the beacon (Section 5.1) is a budget with assumed errors, not a pointing analysis of a real terminal. The alignment between the beacon emitter and the attitude reference of the transmitting satellite, and between the dome and the narrow-field channel of the receiving one, is assumed calibrated. The readout bandwidth and the on-detector block-coincidence logic are assumed, not designed.
\item The statistical uncertainty of the Monte Carlo results is below 1\% (Section 4.9) and is much smaller than the uncertainty that comes from the assumptions: the sensitivity analysis of the Appendix changes the results by tens of percent or by factors of several. The SEs do not include the model assumptions and are not a confidence interval on a real device.
\item The literature and patent comparison relies exclusively on public papers and openly available documents; it is a technical positioning, not a legal novelty or freedom-to-operate analysis.
\end{itemize}

\textbf{Validation needed:} an optical bench with an InGaAs or SPAD pixel array, measurement of the DCR and of the temperature effect, irradiation tests of an array, and a test of the time-stamp chain with simulated attitude.

\section{Conclusions and potential next steps}

In simulation, a dome receiver with an event-driven InGaAs hybrid detector can detect a pulsed beacon between satellites of an Earth constellation in 12 \textmu{}s to 2 ms and, with 20 pulses (about 0.8 ms), estimate the AoA to about 0.13--0.23 mrad with the binary estimator, below the 0.3 mrad requirement for handover to the narrow channel. Compared with published scan-based acquisition, the search phase (light time plus detection) is shorter by a factor of about $10^{2}$ to $10^{5}$; end to end, with pointing and loop-closure times that I assume, it is 0.16--1.5 s for $S$ = 0.15--1.5 s (median case, 10 mm), comparable to the 0.9 s reported for a terminal with calibrated open-loop pointing [36], but without requiring that pointing at the receiver, at the price of a beacon pointed at it to about 1.25 mrad (1$\sigma$ per axis for a 5 mrad beacon; Sections 2.4 and 5.1). In the model, the result holds under the systematic errors considered and, with attitude jitter above 0.3 mrad, only with per-hit correction and a synchronisation of about 0.5 to 38 \textmu{}s depending on the jitter. All values come from simulation with declared assumptions, not from measurements; with 6--9 mm, which is what the optics allow, the far-case detection time rises to 4.3--116 ms, and an optical crosstalk of 1\% raises it to 0.78 s with 10 mm.

Feasibility depends on cooling to about 233 K (radiation) on a 1 Mpixel detector that does not exist today, and on optics that satisfy the \'etendue and filter constraints of Section 3.2. Around the Moon in the model the architecture is feasible at 1000 km altitude, and at 5000 km only for the nearest satellites; a first-order estimate of the radiation there gives a life set by solar particle events, with about 4 g/cm\textsuperscript{2} of aluminium-equivalent shielding needed for a margin against the 90--95\% event and about 7 g/cm\textsuperscript{2} or more against the 99\% one. In the Earth-Moon link the beacon must have 200--500 \textmu{}rad, and the geometry of the Moon and Earth discs limits wide-field acquisition to high orbits.

The simulations are consistent with the hypothesis that coincidence, time stamps and centroiding, borrowed from HEP pixel detectors, transfer to this problem; they do not test it, which requires measurement on a real array.

Following this proof of concept, the following work should be carried out as potential future steps:

\begin{itemize}[leftmargin=*, itemsep=2pt]
\item a transport calculation, with secondary particles, of the Earth and cislunar environments, and solar particle events in the Earth case;
\item a stray-light model with measured BRDF and ghosts;
\item an optical bench with an InGaAs SPAD pixel array and a test of the time-stamp with simulated attitude;
\item irradiation and annealing tests on an array.
\end{itemize}

\section*{Declarations}

The research idea, the receiver architecture (in particular the transfer of the design philosophy of high-energy-physics pixel detectors to the acquisition of inter-satellite links), the choice of the scenarios (an Earth constellation, a lunar constellation and the Earth-Moon link) and of their orbital geometries, and the interpretation of the results are my own. I used Claude Sonnet 5.5 (Anthropic) to help write and debug the simulation code, perform some intermediate calculations and edit the language. I defined the models and the simulations, checked all code, calculations and text, and take full responsibility for the content.

\appendix

\section{Appendix: parameters and assumptions}

\begin{table}[H]
\centering
\footnotesize
\caption{Parameters and assumptions.}
\begin{tabularx}{\linewidth}{>{\raggedright\arraybackslash}X >{\raggedright\arraybackslash}X >{\raggedright\arraybackslash}X}
\toprule
\textbf{Quantity} & \textbf{Value} & \textbf{Origin} \\
\midrule
Beacon average power & 0.5 W (2--10 W in Earth-Moon) & assumption \\
Pulse energy, duration & maximum 20 \textmu{}J, 8 ns & assumption \\
Optical efficiency \ensuremath{\times} PDE & 0.5 \ensuremath{\times} 0.30 & assumption (best published array: 15\% at 295 K [20]) \\
Receiver aperture & 10 mm (30 mm, and up to 100 mm in the Earth-Moon link, as hypothetical references) & assumption; 6--9 mm with f/1.4--f/1 optics (Section 3.2) \\
Pixels, pitch & 1 Mpixel, 25 \textmu{}m & assumption \\
Nominal DCR & 10\textsuperscript{4}/s per pixel & assumption, requires cooling \\
Dead time & 1 \textmu{}s & assumption \\
Target FAR & 0.01/s over the whole frame & assumption \\
Block capture fraction, spot size, AoA requirement & C = 0.7; $\sigma_{PSF}$ = 1.3 mrad; $\sigma_{target}$ = 0.3 mrad & assumption \\
Linear mode & G=20, F=9.2, $\sigma$e=50 e\textsuperscript{$-$}, 5$\sigma$ threshold, QE 0.8 & assumption \\
Systematic errors, realistic / pessimistic & gain 3\% / 10\%, crosstalk 1\% / 5\%, dead pixels 0.5\% / 2\%, jitter 0.1 / 0.3 mrad & assumption \\
Radiation damage & K from Hamamatsu PIN, 105 MeV & published measurement \\
Avalanche trigger probability & 0.30 & assumption \\
Dark-count halving & every 8 K & assumption \\
Hot pixels & lognormal, $\sigma$=1.5; masked above 20\ensuremath{\times} & assumption \\
End-of-life criterion & more than 10\% of pixels masked, or mean DCR above 3\ensuremath{\times} nominal & assumption \\
Annual fluences & 6.9\ensuremath{\times}10\textsuperscript{8}, 1.4\ensuremath{\times}10\textsuperscript{9}, 3.0\ensuremath{\times}10\textsuperscript{9} p/cm\textsuperscript{2} (mean); 1.4\ensuremath{\times}10\textsuperscript{9}, 2.8\ensuremath{\times}10\textsuperscript{9}, 6.3\ensuremath{\times}10\textsuperscript{9} (95th percentile) & AP9 (SPENVIS, IRENE 1.50), 550 km, 53\textdegree{}, shielding 4, 1 and 0.25 g/cm\textsuperscript{2} Al \\
Cislunar fluences & Galactic cosmic rays (400 / 1100 MV, factor 1.5 for ions); solar particles for a 2-year mission at 90 / 95 / 99\% & published models [26], [27]; ESP-PSYCHIC spectra generated by me with the SPENVIS tool; the factor for ions is my assumption \\
Thermal & TEC 12 W, max $\Delta$T 65 K, Carnot fraction 0.13; radiator and shield variable & assumption \\
Attitude & noise 0.05 mrad, bias 0.05 mrad & assumption \\
\bottomrule
\end{tabularx}
\end{table}

\subsection{Rationale and sensitivity of each assumption}

Table 5 lists the assumptions; Table 6 gives, for each one, the reason for the value and what happens to the results when it changes. Each row varies one parameter at a time, with the others at their nominal value. Detection times are for a 5 mrad beacon, DCR 10\textsuperscript{4}/s, in the near, median and far cases. "Impact" is low if the result changes by less than about 10\%, moderate up to a factor 2, high above that. For a few assumptions I did not run a sensitivity, and the table says so.

\begingroup\footnotesize\setlength{\tabcolsep}{3pt}
\begin{longtable}{>{\raggedright\arraybackslash}p{0.19\linewidth} >{\raggedright\arraybackslash}p{0.27\linewidth} >{\raggedright\arraybackslash}p{0.38\linewidth} >{\raggedright\arraybackslash}p{0.10\linewidth}}
\caption{Basis, sensitivity and impact of each assumption.}\\
\toprule
\textbf{Assumption and value} & \textbf{Reason for the value} & \textbf{Sensitivity (computed)} & \textbf{Impact} \\
\midrule
\endfirsthead
\toprule
\textbf{Assumption and value} & \textbf{Reason for the value} & \textbf{Sensitivity (computed)} & \textbf{Impact} \\
\midrule
\endhead
Receiver aperture 10 mm & Round number slightly above the 6--9 mm that a single lens per pixel group on this detector allows (Section 3.2); 30 mm is not realisable with a single lens per pixel group and is kept as a reference. & Near, median and far detection times at 5 mrad: 6 mm 33 \textmu{}s / 5.1 ms / 116 ms; 9 mm 14 \textmu{}s / 0.36 ms / 4.3 ms; 10 mm 12 \textmu{}s / 0.20 ms / 2.0 ms; 30 mm 1.3 / 14 / 31 \textmu{}s. & High \\
Pointing of the narrow channel 0.1--1 s and loop closure 0.05--0.5 s & Order-of-magnitude estimates of mine for steps common to the scan-based approach and to mine, not simulated (Section 2.4): the lower end of the pointing is the 16 ms per step of a fast steering mirror [36], the upper end is 5--8.5 mrad/s over a few milliradians [37]; loop closure comes from sampling at 1 kHz [37] and a transition of less than one second [38]. The coarse slew is excluded. & End-to-end time 0.16, 1.5 and 10 s for $S$ = 0.15, 1.5 and 10 s, and an advantage over the scan-based references from 1.2 to 2600 (Table 2); with $S$ = 0 the receiver would take 3.4--132 ms. & High for the end-to-end comparison, none for the search phase \\
Beacon power 0.5 W & My choice of a low-power transmitter; not derived from a link requirement. & Detection time is inversely proportional to power: with 10 mm, 0.25 W gives 24 \textmu{}s / 0.41 ms / 4.0 ms and 1 W gives 5.8 / 102 \textmu{}s / 1.0 ms; with 30 mm, 2.6 / 29 / 62 \textmu{}s and 1.0 / 7.2 / 16 \textmu{}s. & High \\
Pulse energy up to 20 \textmu{}J, 8 ns (peak 2.5 kW) & Sets the longest pulse period, 40 \textmu{}s. My first assumption of 2 ns (10 kW peak) was too high for a compact space amplifier, so I use 8 ns (2.5 kW). A published space-oriented 1.5 \textmu{}m fibre-amplifier transmitter reports 1 kW peak, 6 W average and 0.5--8 ns pulses (ILRC 2016), so 2.5 kW is about 2.5 times the level I found demonstrated in a compact space configuration. Ground-based fibre amplifiers for lidar in the 1.5 \textmu{}m band already reach larger pulse energies (up to 1.8\,mJ at 2.5\,kHz, with longer pulses [25]), so I assume that a space-qualified transmitter with 2.5 kW peak will exist; this is a scenario assumption, and the 5 \textmu{}J case in the next column is the level demonstrated in space. Eye safety: the ICNIRP 2013 corneal limit at 1500--1800\,nm is 10\,kJ/m\textsuperscript{2} for any pulse from 1\,ns to 10\,s [21], so a 20 \textmu{}J pulse over a 1\,cm aperture is four orders of magnitude below it. The average power is the tighter constraint (1\,kW/m\textsuperscript{2} over 10\,s), which a 0.5\,W, 5\,mrad beam meets beyond about 5\,m, and 2--10\,W at 200\,\textmu{}rad beyond about 250--560\,m. At the exit aperture the beacon is not Class 1 (about 10\,mW through 3.5\,mm [22], known to me only from public excerpts, not the full Ed.\ 3 text); the formal classification is a design task outside this paper. & 5 \textmu{}J (about 1 kW peak at 5 ns, the level demonstrated in the literature I found): 10 mm gives 12 \textmu{}s / 5.3 ms / 149 ms; 30 mm gives 1.3 / 14.5 / 45 \textmu{}s. The pulse duration enters only through the coincidence window, which cannot be shorter than the pulse: with 2 ns instead of 8 ns the 10 mm far case goes from 2.01 to 1.97 ms and the 30 mm results do not change. 100 \textmu{}J: 10 mm gives 12 \textmu{}s / 0.13 ms / 0.29 ms. 20 \textmu{}J with 20 ns (1 kW peak): 10 mm gives 12 \textmu{}s / 0.50 ms / 9.5 ms (the window grows to 20 ns); 30 mm gives 1.3 / 14.3 / 31.7 \textmu{}s. & High (10 mm), moderate (30 mm) \\
Optical efficiency 0.5, PDE 0.30 & The efficiency is a typical value that I assume; the PDE is twice the best published array that I found (15\% at 295 K [20]) and is optimistic. & Time scales as the inverse of their product. With PDE 0.15 (or efficiency 0.25): 10 mm gives 24 \textmu{}s / 1.6 ms / 29 ms; 30 mm gives 2.7 / 29 / 63 \textmu{}s. With a product 1.5 times larger: 10 mm gives 7.8 / 100 / 550 \textmu{}s; 30 mm gives 1.0 / 9.8 / 21 \textmu{}s. & High \\
Block capture fraction C = 0.7 & Average fraction of the signal in the sliding block, an assumption. & 10 mm: C = 0.5 gives 16 \textmu{}s / 0.51 ms / 6.9 ms and C = 0.9 gives 9.0 / 119 \textmu{}s / 0.88 ms; 30 mm: 1.8 / 21 / 44 \textmu{}s and 1.0 / 11 / 24 \textmu{}s. & High (10 mm), moderate (30 mm) \\
Spot size $\sigma_{\mathrm{PSF}}$ = 1.3 mrad & Chosen to spread the spot over a cluster for the centroid. It sets the counts needed for the angle, ($\sigma_{\mathrm{PSF}}$/$\sigma_{\mathrm{target}}$)\textsuperscript{2}. & 10 mm: 0.65 mrad gives 7.8 \textmu{}s / 0.20 ms / 2.0 ms and 2.6 mrad gives 46 \textmu{}s / 0.52 ms / 2.0 ms; 30 mm: 1.0 / 9.5 / 21 \textmu{}s and 5.1 / 57 / 123 \textmu{}s. & Moderate \\
1 Mpixel, 25 \textmu{}m pitch & Gives a pixel of 2.92 mrad over 8.5 sr. The array is larger than any published one, and the pitch fixes the optics (Section 3.2). & In the false-alarm budget, 0.25 to 4 Mpixel does not change the times (the far case at 10 mm is 2.01 ms in both cases). The effect on the optics is not covered by this number. & Low in the budget, high on the optics \\
DCR 10\textsuperscript{4}/s & Design assumption, with the plausibility check in Section 4.3. & 10 mm, at 10\textsuperscript{3} / 10\textsuperscript{4} / 10\textsuperscript{5}/s: near 11.6 / 11.8 / 13 \textmu{}s, median 0.13 / 0.20 / 1.6 ms, far 0.58 / 2.0 / 65 ms. 30 mm: less than 15\% over the same range. & High (10 mm), low (30 mm) \\
Dead time 1 \textmu{}s & Assumed longer than the pulse (8 ns) and the coincidence window (up to 50 ns), so a pixel gives one hit per pulse. It equals the shortest pulse period searched (1 \textmu{}s). Published free-running InGaAs/InP SPADs typically use 1--10 \textmu{}s (relaxation up to 50 \textmu{}s), so 1 \textmu{}s is at the optimistic end. & Pulse period forced to be at least the dead time: with 10 mm, 1 / 2 / 5 / 10 \textmu{}s give 11.8 / 11.8 / 11.8 / 11.9 \textmu{}s in the near case, and the median and far cases (0.20 and 2.0 ms) do not change; with 30 mm the near case becomes 1.3 / 2.0 / 5.1 / 10.3 \textmu{}s, and the median and far cases (14 and 31 \textmu{}s) do not change. The model treats a pixel as firing once per pulse and does not model afterpulsing from the dead time. & Low \\
Target false-alarm rate 0.01/s & One false alarm per 100 s over the whole frame; no operational requirement exists for it. & With 10 mm, 1/s changes the far case from 2.01 to 1.96 ms, and 10\textsuperscript{$-$4}/s gives 0.50 ms (median) and 9.6 ms (far); with 30 mm, 10\textsuperscript{$-$4}/s and 1/s change the times by less than 1\%. & High at 10\textsuperscript{$-$4}/s (10 mm), low (30 mm) \\
Linear mode: G = 20, F = 9.2, $\sigma_{e}$ = 50 e\textsuperscript{$-$}, 5$\sigma$ threshold, QE 0.8 & Assumed values for an APD chain of the kind that the ToT estimator needs. Not taken from a specific device. & Corrected ToT error, median case: $\sigma_{e}$ = 25 / 50 / 100 / 200 e\textsuperscript{$-$} gives 0.38 / 0.74 / 1.08 / 0.86 mrad; G = 10 / 20 / 40 gives 1.09 / 0.74 / 0.38; F = 4 / 9.2 / 15 gives 1.03 / 0.73 / 0.61. With 10 mm the signal is a few counts per pixel and the ToT estimate depends on the threshold in a way that is not monotonic (for example at 200 e\textsuperscript{$-$} and for F), which I did not investigate; I read these values only as ``above 0.3 mrad except for the least noisy chains''. The binary estimator does not depend on them (0.24 mrad). QE not varied. Setup of these runs: spot of 0.8 pixel, 0.3 mrad jitter at 100 Hz, 10 ms latency, 3\ensuremath{\times}10\textsuperscript{4} trials, SE 0.005 mrad or less. & High for ToT \\
Damage constant K & Measured on 1 mm InGaAs PIN photodiodes at 105 MeV [10] and scaled to a 25 \textmu{}m pixel; applying it to a SPAD pixel and to other proton energies is an extrapolation (proton damage of InGaAs/InP SPADs at 3 MeV is reported in [10] and was not used to set K). & K \ensuremath{\times} 0.5 / \ensuremath{\times} 2: life at 233 K and 1.5\ensuremath{\times}10\textsuperscript{9} p/cm\textsuperscript{2} per year is 16 / 4.0 years (nominal 8.0). & High \\
Avalanche trigger probability 0.30 & Assumed. & 0.1 / 0.6: 24 / 4.0 years. & High \\
Dark-count halving every 8 K & Assumed from the common behaviour of InGaAs dark current; not measured on this array. & 6 K / 10 K: 49 / 2.7 years. & High \\
Hot pixels: lognormal $\sigma$ = 1.5, mask above 20\ensuremath{\times} & Assumed distribution and threshold. & $\sigma$ = 1.0 / 2.0: 6.0 / 15 years. Threshold 10\ensuremath{\times} / 50\ensuremath{\times}: 11.7 / 6.3 years. & Moderate \\
Annual fluences 6.9\ensuremath{\times}10\textsuperscript{8} to 6.3\ensuremath{\times}10\textsuperscript{9} p/cm\textsuperscript{2} & AP9 model [30], 550 km and 53\textdegree{}, three shieldings, run in SPENVIS (not by me; checked to 10\%). Solar events are not included. & Life at 233 K: 17 / 8.6 / 4.0 years (mean) and 8.8 / 4.3 / 1.9 years (95th percentile) for the three shieldings (the two cases at about 1.4\ensuremath{\times}10\textsuperscript{9} p/cm\textsuperscript{2}, 8.6 and 8.8 years, differ because the fluences are rounded here). A different model could change them by a factor not estimated here. & High \\
Cislunar environment (first-order estimate) & Public models: cosmic-ray spectrum of [26] with force-field modulation, ESP-PSYCHIC spectra [27] generated with SPENVIS, damage function of silicon with a plateau at high energy, no secondaries. & Open space, 1 g/cm\textsuperscript{2}, solar minimum: without the plateau of the damage function the cosmic-ray fluence is 18\% lower (protons only, 7.0 against 8.6\ensuremath{\times}10\textsuperscript{7} p/cm\textsuperscript{2} per year); a factor for ions of 1.3 or 2.0 instead of 1.5 gives 1.1 or 1.7\ensuremath{\times}10\textsuperscript{8}. Solar particles at 233 K: life with 0.25 / 1 / 4 g/cm\textsuperscript{2} of 0 / 0 / 69 years (90\%) and 0 / 0 / 0 years (99\%), cosmic-ray minimum. & High (solar events), low (cosmic rays) \\
Thermal: TEC 12 W, maximum $\Delta$T 65 K, Carnot fraction 0.13 & Order of magnitude that I assume. A commercial two-stage module (Marlow MS2 series, datasheet seen on a distributor site) gives about 82 K maximum $\Delta$T and about 10 W Qmax with a 300 K hot side, so 65 K and 12 W is the same order. Real modules depend on the hot-side temperature and degrade at the cold end. & One parameter at a time, with 5 W of detector dissipation and a 50\% sun shield, at the worst beta. With a 0.10 m\textsuperscript{2} radiator, 233 K is held under every variation except the TEC capacity: with 6 W instead of 12 W the TEC saturates and the detector is not held at 233 K (the solver reaches its bound, so the temperature returned is not meaningful), while 24 W changes nothing. Maximum $\Delta$T 50 / 80 K, Carnot fraction 0.07 / 0.26, parasitic conductance 0.015 / 0.06 W/K, effective emissivity 0.15 / 0.6, bus 283 / 303 K, Earth IR 150 / 300 W/m\textsuperscript{2}, radiator emissivity 0.75 / 0.92 and parasitic flux 7.5 / 30 W/m\textsuperscript{2} do not move it. With the 0.05 m\textsuperscript{2} radiator, which is already at 248 K, the effective temperature becomes 263 / 237 K for a Carnot fraction of 0.07 / 0.26, 243 / 255 K for a parasitic conductance of 0.015 / 0.06 W/K and 254 / 244 K for a radiator emissivity of 0.75 / 0.92; with the halving every 8 K, a 15 K shift multiplies the damage-induced dark count by about 3.7. & High (TEC capacity), moderate (0.05 m\textsuperscript{2}) \\
Systematic errors: gain 3\%, crosstalk 1\%, dead pixels 0.5\%, jitter 0.1 mrad (realistic) & Typical values that I assume after calibration; not taken from a specific device. & Median case with 10 mm, binary / ToT, in mrad: no systematics 0.173 / 0.332; gain 5\% / 10\% / 20\%: 0.174 / 0.338, 0.180 / 0.354, 0.205 / 0.414; crosstalk 1--5\%: 0.175--0.185 / 0.343--0.394; dead pixels 0.5\% / 2\%: 0.184 / 0.343, 0.215 / 0.375; jitter 0.1 / 0.3 / 1 mrad: 0.173 / 0.330, 0.182 / 0.332, 0.281 / 0.435; realistic 0.187 / 0.356, pessimistic 0.235 / 0.453 (3\ensuremath{\times}10\textsuperscript{4} trials, SE 0.003 mrad or less). & Moderate (binary); ToT already above 0.3 mrad \\
AoA requirement 0.3 mrad & Assumed accuracy for the handover to the narrow channel; with $\sigma_{PSF}$ = 1.3 mrad it fixes the counts needed, ($\sigma_{PSF}$/$\sigma_{target}$)\textsuperscript{2} \ensuremath{\approx} 19. & 0.15 mrad (about 75 counts): 10 mm gives 46 \textmu{}s / 0.5 ms / 2.0 ms; 30 mm gives 5 / 57 / 123 \textmu{}s. & Moderate (near and median), low (far) \\
Attitude measurement: 0.05 mrad noise, 0.05 mrad bias & Assumed order of magnitude for a star tracker with gyro. & Noise and bias 0.02 / 0.05 / 0.1 / 0.2 mrad: corrected AoA error 0.23 / 0.24 / 0.25 / 0.31 mrad for 0.3 mrad jitter (binary, median case, spot of 0.8 pixel), so the last value exceeds 0.3 mrad (SE 0.001 mrad or less, 3\ensuremath{\times}10\textsuperscript{4} trials). The uncorrected error does not change. & High above 0.1 mrad \\
\bottomrule
\end{longtable}
\endgroup

\end{document}